\documentclass[journal]{IEEEtran}
\usepackage{array}
\usepackage{epsfig}
\usepackage{cite}
\usepackage{epstopdf}
\usepackage{amsfonts,amsmath,amsthm,amssymb,balance}
\usepackage{multirow,balance}
\usepackage{bm}
\usepackage[usenames,dvipsnames]{color}
\usepackage{colortbl}
\usepackage{float}
\usepackage[ruled,vlined,linesnumbered]{algorithm2e}
\usepackage{graphicx}
\usepackage{subfigure}
\usepackage{thmtools}
\definecolor{mygray}{gray}{.9}

\graphicspath{{figure/}}
\newcolumntype{C}[1]{>{\PreserveBackslash\centering}p{#1}}
\newcolumntype{R}[1]{>{\PreserveBackslash\raggedleft}p{#1}}
\newcolumntype{L}[1]{>{\PreserveBackslash\raggedright}p{#1}}
\newcommand{\AlgoResetCount}{\renewcommand{\@ResetCounterIfNeeded}{\setcounter{AlgoLine}{0}}}
\newcommand{\AlgoNoResetCount}{\renewcommand{\@ResetCounterIfNeeded}{}}
\newcounter{AlgoSavedLineCount}

\theoremstyle{definition}

\newtheorem{proposition}{Proposition}
\newtheorem{corollary}{Corollary}
\newtheorem{lemma}{Lemma}
\newtheorem{remark}{Remark}

\begin{document}

\title{Physical Layer Security Assisted Computation Offloading in Intelligently Connected Vehicle Networks}

\author{\IEEEauthorblockN{Yiliang Liu, Wei Wang,~\IEEEmembership{Member,~IEEE}, $\mbox{Hsiao-Hwa~Chen}${$^{^\dagger}$},~\IEEEmembership{Fellow,~IEEE}, \\
Feng Lyu,~\IEEEmembership{Member,~IEEE}, Liangmin Wang,~\IEEEmembership{Member,~IEEE}, \\
Weixiao Meng,~\IEEEmembership{Senior Member,~IEEE}, and Xuemin (Sherman) Shen,~\IEEEmembership{Fellow,~IEEE}}

\thanks{Y. Liu  (email: {\tt alanliuyiliang@gmail.com}) and W. Meng (email: {\tt wxmeng@hit.edu.cn}) are with the School of Electronics and Information Engineering, Harbin Institute of Technology, Harbin 150001, China. W. Wang (email: {\tt wei\_wang@nuaa.edu.cn}) is with the College of Electronic and Information Engineering, Nanjing University of Aeronautics and Astronautics, Nanjing 210016, China. H.-H. Chen (email: {\tt hshwchen@ieee.org}) (the corresponding author) is with the Department of Engineering Science, National Cheng Kung University, Tainan 70101, Taiwan. F. Lyu (email: {\tt fenglyu@csu.edu.cn}) is with the School of Computer Science and Engineering, Central South University, Changsha 410083, China. X. Shen (email: {\tt sshen@uwaterloo.ca}) is with the Department of Electrical and Computer Engineering, University of Waterloo, Waterloo N2L 3G1, Canada. L. Wang (email: {\tt wanglm@ujs.edu.cn}) is with the School of Computer Science and Communication Engineering, Jiangsu University, Zhenjiang 212000, China.}

\thanks{This work was supported in part by National Natural Science Foundation of China (U1764263 and 61671186), Taiwan Ministry of Science and Technology (109-2221-E-006-175-MY3 and 109-2221-E-006-182-MY3), and Natural Sciences and Engineering Research Council (NSERC) of Canada.}

\thanks{Manuscript received January 4, 2020; revised on August 25, 2020 and November 26, 2020; accepted on January 1, 2020.}

}

\markboth{IEEE Transactions on Wireless Communications, VOL. XX, MONTH YY, YEAR 2021}{Liu \MakeLowercase{\textit{et al.}}: Physical Layer Security Assisted\ldots}

\maketitle

\begin{abstract}
In this paper, we propose a secure computation offloading scheme (SCOS) in intelligently connected vehicle (ICV) networks, aiming to minimize overall latency of computing via offloading part of computational tasks to nearby servers in small cell base stations (SBSs), while securing the information delivered during offloading and feedback phases via physical layer security. Existing computation offloading schemes usually neglected time-varying characteristics of channels and their corresponding secrecy rates, resulting in an inappropriate task partition ratio and a large secrecy outage probability. To address these issues, we utilize an ergodic secrecy rate to determine how many tasks are offloaded to the edge, where ergodic secrecy rate represents the average secrecy rate over all realizations in a time-varying wireless channel. Adaptive wiretap code rates are proposed with a secrecy outage constraint to match time-varying wireless channels. In addition, the proposed secure beamforming and artificial noise (AN) schemes can improve the ergodic secrecy rates of uplink and downlink channels even without eavesdropper channel state information (CSI). Numerical results demonstrate that the proposed schemes have a shorter system delay than the strategies neglecting time-varying characteristics.
\end{abstract}

\IEEEpeerreviewmaketitle

\vspace{0.1in}
\section{INTRODUCTION}
The advancement in sensors, cloud computing, artificial intelligence (AI), and 5G technologies has pushed the evolution of traditional vehicle-to-vehicle (V2V) networks toward intelligently connected vehicle (ICV) networks\cite{Zhang2018}. Compared to safety or value-added services in V2V networks supported by IEEE 802.11p \cite{reliable-V2V}, 5G-enabled ICVs equipped with advanced computing modules can realize autonomous driving, driver-supervised driving, cooperative driving, high definition and three-dimensional (3D) map services, on-board working and entertainment, augmented reality (AR), etc.\cite{Zhou2017,Yang2018Intelligentconnectedvehicles, Mei2018}. In addition to local computing, enabled by mobile edge computing (MEC) technologies, ICVs can offload part of computational tasks to nearby servers in SBSs to cut down vehicle cost. For example, in Tesla store\cite{Tesla}, consumers should pay an extra 8,000 $\sim$ 10,000 US dollar for full self-driving hardware (Nvidia drive PX 2 platform), whose price could probably be reduced substantially if dedicated MEC services are available\cite{Zhang2018, Mao2017SurveyMobileEdge}. Also, MEC can reduce computing and network latency when on-board computing capacity is insufficient\cite{HuEarlyAccess2019, Cheng2019,Wang2016MobileEdgeComputing,Chen2016EfficientMultiUser,Yang2018SmallCellAssisted, Xu2019a, Bai2019}. 

MEC faces many security threats due to information exchanges between ICVs and SBSs. In order to guarantee information confidentiality, symmetric key agreements with authentication processes were proposed in the literature \cite{Yan2016,Brecht2018SecurityCredentialManagement, Zhou2015SecurePrivacyPreserving}. However, secret key establishment is controlled by centralized parties of cryptography management, such as authentication, authorization, and accounting (AAA) servers defined in the 3GPP standard, which requires ubiquitously available trusted third-party and tamper-proof devices \cite{Zhou2013,Zhang2018HealthDepEfficientSecure, Wazid2020}, and therefore it may be not suitable for highly dynamic ICV networks. As an important security mechanism, physical layer security is capable to achieve confidential information transmission by exploring random characteristics of wireless medium, and is implemented via physical layer technologies (such as encoding, decoding, and resource allocation, etc.) without cryptography operations \cite{Wang2018PhysicalLayerSecurity}. 

The issue of high mobility in vehicular networks should not be ignored for improving the performance of computation offloading schemes for vehicles. Task partition techniques divide computation tasks and determine which part of computation tasks to be offloaded, which is conducted before MEC computation. The task partition requires satisfactory secrecy rates in uplink and downlink, and uses these rates to balance local and edge computations. Previous computation offloading schemes assumed that these rates are time-invariant during offloading duration \cite{HuEarlyAccess2019, Cheng2019, Wang2016MobileEdgeComputing, Chen2016EfficientMultiUser,Yang2018SmallCellAssisted, Xu2019a, Bai2019}. However, the coherence time of vehicular channels is short due to high mobility (usually less than 1 ms), which means that uplink and downlink channels change rapidly, and thus data transmission in ICV networks suffers distortions due to varying wireless channels. In this case, task partition with a time-invariant rate may lead to a poor offloading performance when the rates of data transmission must be made different due to variations of wireless channels. Besides, high mobility also brings in more challenges in data transmissions. Fixed wiretap code rates\footnote{The wiretap code was introduced first by Wyner, which is a general code scheme to ensure that a coding rate (defined as a wiretap code rate) below the secrecy capacity can achieve both reliability and information-theoretic security \cite{Zhou2011RethinkingSecrecyOutage}. The secrecy rate can be viewed as an inherent property of a given communication system, while the wiretap code rate is controlled by encoders. This paper uses Wyner's coding that is non-structured random codes based on cosets. Many efforts were dedicated to find practical codes based on Wyner's theory \cite{Harrison2019, Oggier2016, Mahdavifar2011}.} without considering time-varying secrecy rate may cause a large secrecy outage. Note that even we can keep a secrecy outage probability below a threshold, we still can not avoid a secrecy outage completely because the CSIs of eavesdroppers are unavailable during the entire transmission and computation phases. The unknown CSI of eavesdroppers is a typical issue and should not be ignored in physical layer security assisted offloading schemes. The other issue is a long latency caused by small secrecy rates. Existing computation offloading schemes use wiretap coding with single antenna technologies to achieve physical layer security \cite{Yang2018SmallCellAssisted, Xu2019a, Bai2019}. When eavesdroppers have better channel conditions or more antennas, the secrecy rate can be very low and even equals to zero, causing a long delay if using a channel with a small secrecy rate to transmit data. 

In order to address the aforementioned issues, we formulate a latency minimization problem with respect to a computation task partition ratio, where an ICV network uses ergodic secrecy rates of uplink and downlink for computation task partition. Although these time-varying CSIs and their secrecy rates are unavailable in the task partition phase as transmissions have not yet occurred in this phase, the ICV can calculate the ergodic secrecy rates of uplink and downlink by integral equations, considering that the CSIs are integral variables. In the follow-up transmission phases, the main CSIs between ICV and SBS in each transmission burst can be estimated, such that an adaptive wiretap code rate with secrecy outage constraints can be identified according to the main CSIs to avoid a large secrecy outage. In addition, SBS-assisted jamming, multi-antenna beamforming, and artificial noise (AN) technologies are used to improve secrecy rates, which do not require CSIs of eavesdroppers. The main contributions of this work are summarized as follows.
\begin{enumerate}
\item First, we formulate a secure computation offloading model of an ICV network, where joint AN-assisted beamforming and wiretap coding schemes are used to prevent a multi-antenna eavesdropper from wiretapping uplink and downlink information between the ICV and MEC server. Due to the latency requirement of computation services, the optimization problem is to minimize the total latency of transmission and computation. 
\item In computation task partition phases, the exact expressions of uplink and downlink ergodic secrecy rates are deduced for task partition. The closed-form expressions of the lower bounds of these ergodic secrecy rates are given also to reduce complexity. In addition, computation task partition is solved in a closed-form.
\item To improve the secrecy rates of uplink, SBS receives uplink data via maximum ratio combining (MRC), while transmitting AN signals to confuse an eavesdropper without any self-interference at SBS. In downlink, SBS uses maximum ratio transmission (MRT) for message beamforming and sending AN signals in null spaces. To mitigate time-varying effects in wiretap coding, adaptive wiretap code rates are proposed, considering secrecy outage probabilities in both uplink and downlink. 
\end{enumerate}

The remainder of this paper can be outlined as follows. Section \ref{related work} surveys the related works. Section \ref{model} describes the system model and problem formulation, together with the workflow of SCOS given in Section \ref{workflow}. The computation task partition approach, uplink/downlink secure beamforming, and AN technologies are elaborated in Section \ref{task partition}. The schemes enabling adaptive secrecy rates are proposed in Section \ref{coding}. We show simulation results in Section \ref{simulation}, and conclude this paper in Section \ref{conclusion}. 

The major notations are defined as follows. Bold uppercase letters denote matrices and bold lowercase letters denote column vectors. $\mathbf{A}^{\dagger}$ represents the Hermitian transpose of $\mathbf{A}$. $\mathbf{I}_a$ is an identity matrix with its rank $a$. $\mathcal{CN}(\mu,\sigma^2)$ is a complex normal (Gaussian) distribution with mean $\mu$ and variance $\sigma^2$. $(\mathbf{A})^{-1}$ is an inverse function of $\mathbf{A}$. $|\mathbf{x}|$ is the Euclidean norm of $\mathbf{x}$. E$[\cdot]$ is the expectation operator. $\binom{x}{y}$ is the combination between $x$ and $y$ such that $\binom{x}{y}=\frac{x!}{(x-y)!y!}$. $\text{Rank}(\mathbf{A})$ calculates the rank of $\mathbf{A}$. $x!$ is the factorial of $x$. $e\simeq 2.7183$ is a constant. $\text{diag}(\mathbf{x})$ is a diagonal matrix of vector $\mathbf{x}$. An $[a\times(b+c)]$ matrix $[\mathbf{A},\mathbf{B}]$ denotes a combined matrix between an $(a\times b)$ matrix $\mathbf{A}$ and an $(a\times c)$ matrix $\mathbf{B}$.

\vspace{0.1in}
\section{RELATED WORKS}\label{related work}
In this section, we briefly discuss about MEC and traditional AN-based physical layer security technologies considered in SCOS designs.

\subsection{MEC Technologies}
As mentioned in \cite{Mao2017SurveyMobileEdge}, network, communication, computation task partition, and security are the main technical challenges of MEC. Sardellitti \textsl{et al.} focused on the physical layer in network and communication aspects, where energy consumption of all users was minimized via beamforming and computation resources optimization \cite{Sardellitti2015JointOptimizationRadio}. The MEC/cloud access capacity was discussed in \cite{Chen2016EfficientMultiUser}, which maximized the number of users that access to the cloud. Wang \textsl{et al.} proposed a routing algorithm based on deep reinforcement learning, which aimed to minimize routing delay and improve network bandwidth utilization \cite{Wang2019SmartResourceAllocation}. Two classic task partition models were proposed as follows. Miettinen \textsl{et al.} considered a data-based model, where task-input data were bit-wise independent and can be arbitrarily divided into different groups and executed by different entities simultaneously \cite{Miettinen2010Energyefficiencymobile}. Mahmoodi \textsl{et al.} proposed a taskcall graph-based model that decides whether a component completes processing on a mobile or edge server \cite{Mahmoodi2019OptimalJointScheduling}. The taskcall graph has three typical dependency models, i.e., sequential, parallel, and general computation task dependency models. Cheng \textsl{et al.} considered a parallel computation task dependency model and assigned virtual machines (VMs) in an edge server to the tasks that can be executed in parallel \cite{Cheng2019}. 

The ICV computation tasks can be categorized into time-sensitive tasks and non-mission critical tasks. As mentioned in \cite{Yang2018Intelligentconnectedvehicles}, time-sensitive ICV tasks include driver-supervised driving, cooperative driving, and autonomous driving, which can perform driving tasks in certain conditions while a driver should intervene whenever needed according to road conditions and system requests. Computation offloading of MEC is capable of reducing the cost of vehicle computing platforms and accelerate computation processes. Meanwhile, it can provide global information on road environments, and the models in an MEC server are more precise than that of vehicles in some machine learning-assisted applications \cite{Mao2017SurveyMobileEdge}. The non-mission critical ICV tasks usually include on-board working and entertainment. Drivers and passengers can take full advantage of MEC to accelerate the processes and enjoy on-board entertainment as these require a large amount of computation power and storage.

To achieve confidentiality, Yang \textsl{et al.} first considered physical layer security in offloading, and used wiretap coding to protect uplink channels from users to MEC servers in a multi-user multi-server scenario \cite{Yang2018SmallCellAssisted}. Considering the security in terrestrial to air channels, Bai \textsl{et al.} used single-antenna jamming technologies to protect uplink channels between an MEC server and an unmanned aerial vehicles (UAV) against both active and passive eavesdroppers \cite{Bai2019}. Zhou \textsl{et al.} extended jamming technologies further in a scenario with multiple UAVs, where task partition ratio, UAV locations, and transmission power were optimized jointly to maximize the minimum uplink secrecy capacity among multiple UAVs \cite{Zhou2020}. Likewise, Wang \textsl{et al.} considered energy efficiency in secure offloading systems with multiple legitimate users. They optimized radio resource allocation to minimize energy consumption \cite{Wang2020}. The radio resources in the above investigations were assumed to be orthogonal. Wu \textsl{et al.} used secrecy outage-constrained wiretap coding to protect uplink channels in non-orthogonal multiple access (NOMA) assisted computation offloading systems, where an optimal computation task partition ratio was investigated to minimize energy consumption and secrecy outage probabilities \cite{Wu2020}. Multiple antennas technologies were also used to improve the performance of secure offloading systems. He \textsl{et al.} used multiple antennas of SBS to generate AN signals to impede eavesdropping and leverage full-duplex communication technologies to suppress self-interference \cite{He2020}. 

\subsection{AN-based Physical Layer Security Technologies}
As shown in the literature, physical layer security is achieved by wiretap coding, where messages are reliably transmitted to destinations while the messages are kept confidential to eavesdroppers \cite{Zhou2011RethinkingSecrecyOutage}. In order to make physical layer security more efficient, AN technologies allocate a part of transmission power for generating interference signals, which improve secrecy capacity/rate. The main AN technologies are summarized as follows. 1) Nullspace based AN technology was presented in \cite{Goel2008GuaranteeingSecrecyusing}, which does not require wiretap CSIs. AN signals are transmitted to the nullspace of main channels, such that they do not interfere desired users but only impair eavesdropped channels. 2) Eigenspace based AN technology can identify some appropriate eigenspaces reserved for messages to generate AN signals, instead of selecting nullspaces only for AN signals \cite{Liu2017SecrecyCapacityAnalysis, Liu2019ArtificialNoisyMIMO}. The aforementioned AN technologies belong to linear precoding. 3) Semidefinite programming (SDP) based AN technology needs complex optimization processing to output the optimal beamforming vectors and AN signals, and it offers an optimal secrecy rate \cite{Liao2011QoSBasedTransmit}. The limitation of the SDP based AN technology is the requirement of wiretap CSIs.

\subsection{Discussion}
The existing investigations on the AN methods did not give ergodic secrecy rate expressions of vehicle-to-infrastructure (V2I) links. Besides, high mobility issues should be considered in wiretap coding as discussed in Section I. Thus, in this work, we propose SCOS, which focuses on computation task partition and uplink/downlink AN-aided secure beamforming, followed by adaptive wiretap coding schemes.

\section{SYSTEM MODEL AND PROBLEM FORMULATION}\label{model}
\subsection{System Model}

\begin{figure}[h]
\centering
\includegraphics[width=0.95\linewidth]{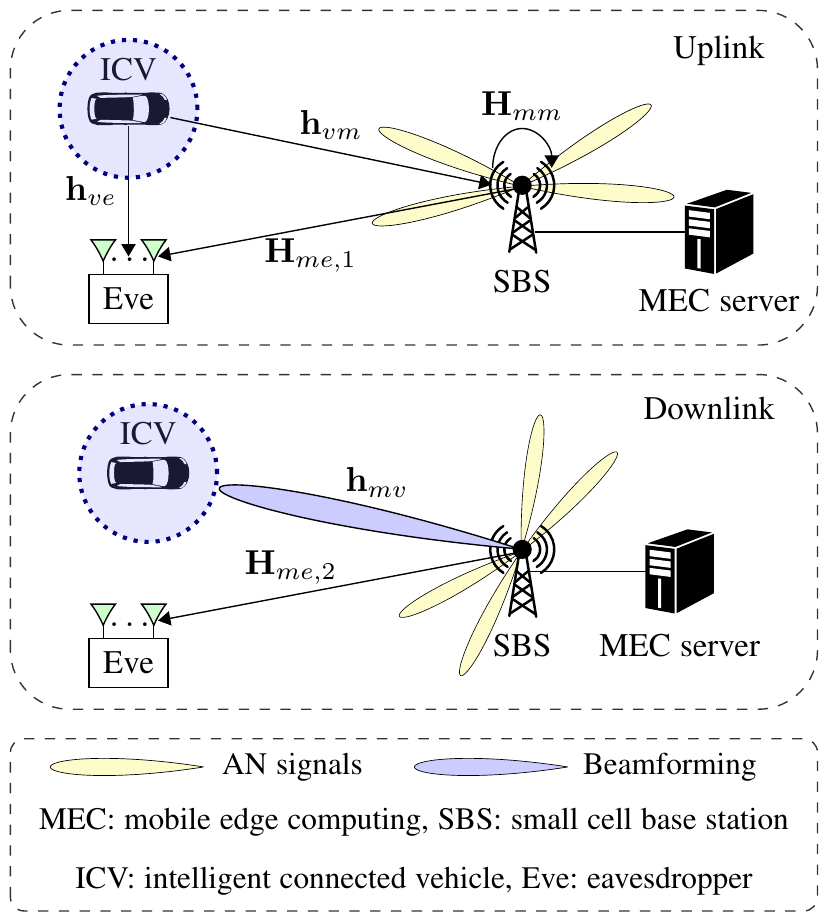}
\caption{A communication model with a multiple antenna eavesdropper. An uplink is to upload a part of computation tasks to a nearby SBS. After the MEC server obtains computation results, a downlink is used to download the computation results to the ICV. In addition, beamforming and AN schemes are illustrated in Section \ref{task partition}.}\label{model_figure}
\end{figure}

Let us consider a secure ICV offloading system, which is equipped with a single antenna, an SBS with $N_m$ antennas, an MEC server, and an eavesdropper (Eve) with $N_e$ antennas, as shown in Fig. \ref{model_figure}. For security purpose, $N_m$ should be larger than $N_e$. If an ICV desires to establish wireless connection for uploading or feedback, it uses Uu interfaces supported by a cellular network with two operational modes on physical layer, i.e., frequency division duplex (FDD) and time division duplex (TDD) \cite{3gpp}. Here, we assume a TDD model because the characteristics of channel reciprocity in TDD avoid CSI feedback overhead. As aforementioned, four entities exist in this system, which are described as follows.
\begin{enumerate}
\item ICV is responsible for computation task partition and transmits a part of tasks to an MEC-assisted SBS confidentially.
\item SBS receives task data while sending AN signals as a cooperative jammer in uplink, and also uses AN-assisted technologies in downlink transmissions to provide feedback to the ICV.
\item MEC server executes computation tasks for the ICV, and is generally installed in an SBS or is located physically close to the SBS, such that the channel between the SBS and the MEC server is assumed to be secure.
\item Eve is a passive eavesdropper that silently receives messages, and thus its CSIs are unavailable.
\end{enumerate}

\subsection{Computation Task Model}
The ICV will perform a computation task with $M$-bits data with the assistance of an MEC server. Let us consider a task that has a full granular data-partition model and can be arbitrarily divided, as discussed in \cite{Wang2016MobileEdgeComputing, Chen2016EfficientMultiUser, Mao2017SurveyMobileEdge}. We define a variable $\eta$ as a ratio of locally processed data to the total amount of data. Hence, in the first phase, $\eta M$ bits will be processed at the ICV and $(1-\eta)M$ bits should be uploaded to SBS. In the second phase, the MEC server of the SBS will complete the computation of $(1-\eta)M$ bits data, and then send the results to the ICV. In both phases, Eve aims to wiretap the information from uplink and downlink channels, and thus physical layer security schemes should be used to protect the uplink and downlink channels. Some investigations ignored the delay of downlink phases when little data need to be transmitted in feedback \cite{Chen2016EfficientMultiUser}. Considering diverse applications in ICVs, such as 3D map \cite{Yang2018Intelligentconnectedvehicles, Mei2018} and remote diagnosis services \cite{Yan2017}, we should not ignore the delay in uplink or downlink phases.

\subsection{Channel Model}\label{channelmodel}
Four channels are considered in the uplink process as shown in Fig. \ref{model_figure}. The uplink channel between the ICV and the SBS is defined as an $N_m\times 1$ vector, i.e., $\mathbf{h}_{vm}$, and a wiretap channel between the ICV and Eve is defined as an $N_e\times 1$ vector, i.e., $\mathbf{h}_{ve}$. As the SBS works as a full-duplex cooperative jammer, the interference channel between the SBS and Eve is defined as an $N_e\times N_m$ matrix, i.e., $\mathbf{H}_{me,1}$. In addition, a self-interference channel incurred from AN signals at the SBS is defined as an $N_m \times N_m$ matrix, i.e., $\mathbf{H}_{mm}$.

Analogously, two channels are considered in the downlink process, as shown in Fig. \ref{model_figure}. The downlink channel between the SBS and the ICV is defined as a $1\times N_m$ vector, i.e., $\mathbf{h}_{mv}$, and the wiretap channel between Eve and the SBS is defined as an $N_e\times N_m$ matrix, i.e., $\mathbf{H}_{me,2}$.

The works in \cite{Patel2005SimulationRayleighfaded} showed that vehicular channels can be simplified by slow-fading Rayleigh models due to the effect of heavily built-up urban environments on radio signals. Also, vehicular CSIs are constant within coherence time, e.g., the coherence time is approximately 2 ms when the speed of vehicles is 10 m/s and the center frequency is 6 GHz. Thus, data delivery is done through multiple wireless channels. According to the aforementioned features, the CSIs of these six channels are assumed as follows.
\begin{enumerate}
\item Main CSIs: In the uplink and downlink, we assume that the CSIs of the main channels $\mathbf{h}_{vm}$, $\mathbf{H}_{mm}$, and $\mathbf{h}_{mv}$ are obtained via pilot-added estimation technologies \cite{3gpp2} in each transmission burst, which do not require the location information of vehicles. In addition, as CSIs of the main channels are time-varying, they are unavailable in the computation task partition phase since offloading is a causal process, and the task partition ratio $\eta$ will be determined based only on the channel distribution information (CDI), assuming that the main CSIs obey Rayleigh fading models. 
\item Wiretap CSIs: Since Eve keeps silent, $\mathbf{h}_{ve}$, $\mathbf{H}_{me,1}$, and $\mathbf{H}_{me,2}$ are unavailable. For simplicity, all of these channels are assumed to obey Rayleigh fading channel models. Since $\mathbf{H}_{me,1}$ and $\mathbf{H}_{me,2}$ are independent and identically distributed, we simplify $\mathbf{H}_{me,1}$ and $\mathbf{H}_{me,2}$ as a complex Gaussian random matrix $\mathbf{H}_{me}$. 
\end{enumerate}

In this paper, we emphasize the secure computation offloading approach and assume that channel estimation is perfect, as what have been assumed in \cite{Bai2019, Zhou2020, Wang2020, Wu2020, He2020, Yang2018SmallCellAssisted}. However, in practice, perfect CSIs may not be available due to channel estimation and feedback errors. The imperfect CSI leads to errors in precoding vectors and impairs wiretap code rates of the proposed scheme, and then further reduces the secrecy rates. For the case of imperfect CSI, many efforts have been independently conducted, such as imperfect CSI modeling, robust precoding, and wiretap coding with CSI uncertainty \cite{He2013,Li2018, Wang2018PhysicalLayerSecurity}, which help to design secure computation offloading schemes under imperfect CSI conditions. We will consider the imperfect CSI in our future works.

\subsection{Problem Formulation}
Here, we first derive a latency expression before formulating a latency minimization problem. In this model, when $\eta M$ bits data are processed in an ICV, the local computing time is $T_v=\eta M/a_v$, and the computing time of the MEC server is $T_m=(1-\eta)M/a_m$, where $a_v$ (bit/s) and $a_m$ (bit/s) are the computing speeds of the ICV and the MEC server, respectively. The transmission delays of uplink and downlink are expressed as $T_u=\beta (1-\eta) M/R_u$ and $T_d=\alpha(1-\eta)M/R_{d}$, respectively, where $\alpha$ accounts for the ratio of output to input bits offloaded to the MEC server, and $\beta$ is the compression ratio of uploaded data as many data types, e.g. images and videos, should be compressed before uploading and decompressed in MEC server before data processing. $R_u$ (bit/s) and $R_d$ (bit/s) are the wiretap code rates of the uplink and downlink channels, respectively. Similar latency models can be seen in \cite{Wang2016MobileEdgeComputing, Munoz2015}. In this case, the latency of this task computation can be formulated as
\begin{flalign}
T_{\text{latency}}(\eta,R_u,R_d)=\max(T_v,T_m+T_u+T_d),
\end{flalign}
where $\max(x,y)$ is the maximum values of $x$ and $y$. Note that $T_{\text{latency}}(\eta,R_u,R_d)$ is a function of $\eta$, $R_u$, and $R_d$. $a_v$ and $a_m$ are fixed computation parameters for the given ICVs and SBSs.

For vehicular tasks where the vehicle has a stringent requirement on the speed of computing feedback, it is preferred to shorten the latency as much as possible, which can be formulated as
\begin{subequations}
\begin{flalign}
\text{P1: } &\min_{\eta}\big\{ T_{\text{latency}}(\eta,R_u,R_d)\big\},  \\
&\text{s.t. }   0 \leq \eta \leq 1. \label{P1C1} 
\end{flalign}
\end{subequations}
Constraint (\ref{P1C1}) specifies the domain of offloading ratio. Due to the non-convexity of the objective function, P1 is a non-convex problem. In addition, it is imprecise to assume that $R_u$ and $R_d$ are time-invariant due to time-varying distortions. We adopt a practical method to solve P1 as follows. In Section \ref{task partition}, we optimize $\eta$ with the ergodic secrecy rates of uplink and downlink, i.e., $\bar{R}_u$ and $\bar{R}_d$. In Section \ref{coding}, we establish uplink and downlink wiretap code rates, considering the secrecy outage probability. 

\subsection{Workflow of SCOS}\label{workflow}

The procedure of SCOS is sketched as follows.
\begin{enumerate} 
\item {\bf Setup:} The ICV handshakes with an SBS and loads the system parameters $\{a_v,a_m,N_m,N_e,P_m,P_v\}$ from the SBS, where $P_{m}$ is the normalized SBS transmission power in both uplink and downlink. $P_{v}$ is the normalized ICV transmission power.
\item {\bf Computation task partition:} The ICV estimates the ergodic secrecy rates of uplink and downlink, i.e., $\bar{R}_u$ and $\bar{R}_d$, with corresponding physical layer security schemes. The ICV picks up a computation task partition based on $\{\bar{R}_u, \bar{R}_d, a_v,a_m\}$, and outputs $\eta^*$. The details are described in Section \ref{task partition}.
\item {\bf Uplink transmission:} The ICV estimates the uplink CSI $\mathbf{h}_{vm}$ within coherence time, then adjusts the wiretap code rate $R_u$ with the constraint of the secrecy outage probability, and sends $\beta (1-\eta^*)M$ data to the SBS at $R_u$. The details are described in Section \ref{upcoding}.
\item {\bf Parallel computing:} The ICV and the SBS execute tasks simultaneously.
\item {\bf Downlink transmission:} The SBS estimates the downlink CSI $\mathbf{h}_{mv}$ within coherence time, then adjusts the wiretap code rate $R_d$ with the constraint of the secrecy outage probability, and sends the results to the ICV at $R_d$. The details are described in Section \ref{docoding}.
\end{enumerate}

In uplink phases, SBS uses a joint MRC and designed AN scheme; while in downlink phases, the SBS uses a joint MRT and nullspace-based AN scheme. We want to mention that SCOS just provides a theoretical guidance to optimize computation task partition and to design proper physical layer security approaches. However, in real-world communication systems, many factors, such as CSI estimation error, signal synchronization, and coding and modulation efficiency, will actually affect the deployment and performance of SCOS. 

\vspace{0.1in}
\section{TASK PARTITION AND AN-ASSISTED PHY-LAYER SECURITY}\label{task partition}
An ICV uses parameters $\{\bar{R}_u, \bar{R}_d, a_v,a_m\}$ for computation task partition. In the setup process, the system parameters $\{a_v,a_m,N_m,N_e,P_m,P_v\}$ are given. Thus, we should first calculate the ergodic secrecy rates of uplink and downlink, i.e., $\bar{R}_u$ and $\bar{R}_d$, which depend on $\{N_m,N_e,P_m,P_v\}$, statistical CSI model (a Rayleigh fading channel), and the corresponding physical layer (PHY-layer) security schemes. The calculations of $\bar{R}_u$ and $\bar{R}_d$ do no require instantaneous CSIs of both main and wiretap channels. 

\subsection{Uplink Ergodic Secrecy Rate $\bar{R}_d$}\label{upprecoding}
The ICV transmits $\beta(1-\eta)M$ data to SBS. The SBS uses an MRC receiver, i.e., $\mathbf{w}_m=\mathbf{h}_{vm}^{\dagger}/|\mathbf{h}_{vm}|$ to gather the signals from different antennas simultaneously, and generates AN signals to confuse Eve. In this case, the received signals at the SBS and Eve are expressed as
\begin{subequations}
\begin{flalign}
&\mathbf{y}_{vm}=\mathbf{h}_{vm}x_u+\mathbf{H}_{mm}\mathbf{G}_u\mathbf{z}_u+\mathbf{n}_{vm}, \\
&\mathbf{y}_{ve}=\mathbf{h}_{ve}x_u+\mathbf{H}_{me}\mathbf{G}_u\mathbf{z}_u+\mathbf{n}_{ve},
\end{flalign}
\end{subequations}
where $x_u$ is the ICV uplink signal encoded by wiretap coding to satisfy $\text{E}(|x_u|^2)=P_{v}$. $\mathbf{n}_{vm}$ and $\mathbf{n}_{ve}$ are AWGN vectors obeying $\mathcal{CN}(\mathbf{0},\sigma_{vm}^2\mathbf{I}_{N_m})$ and $\mathcal{CN}(\mathbf{0},\sigma_{ve}^2\mathbf{I}_{N_e})$, respectively. $\mathbf{G}_u$ is an $N_m\times (N_m-1)$ matrix\footnote{Assume that the number of SBS antennas is larger than that of Eve, i.e., $N_m>N_e$, and Eve has $N_m-1$ antennas. Thus, Eve can not separate the AN signals of SBS from the signals of ICV as the dimension of the spaces for AN signals $\mathbf{G}_u$ is $N_m-1$.}, which lies in the nullspace of $\mathbf{h}_{vm}^{\dagger}\mathbf{H}_{mm}$. $\mathbf{z}_u$ is a complex Gaussian AN signal obeying $\mathcal{CN}(\mathbf{0},\frac{P_m}{N_m-1}\mathbf{I}_{N_m-1})$. 

The processed signals can be formulated as
\begin{flalign}
\mathbf{w}_m\mathbf{y}_{vm}=\mathbf{w}_m\mathbf{h}_{vm}x_u+\mathbf{w}_m\mathbf{n}_{vm}.
\end{flalign}
Although the ICV can obtain $\mathbf{h}_{vm}$ via channel estimation in each uplink transmission burst, it is unavailable in the task partition phase. Thus, we will use the ergodic secrecy rate for task partition. With the proposed MRC and AN schemes, the real ergodic secrecy rate in uplink channel is expressed as
\begin{flalign}\label{escu}
\ddot{R}_{u}&=\text{E}_{\mathbf{h}_{vm},\mathbf{h}_{ve},\mathbf{H}_{me}}[C_{v,u}-C_{e,u}]^{+} \\ 
&\geq \text{E}_{\mathbf{h}_{vm}}[C_{v,u}]-\text{E}_{\mathbf{h}_{vm},\mathbf{h}_{ve},\mathbf{H}_{me}}[C_{e,u}], \label{lbscu}
\end{flalign}
where 
\begin{subequations}
\begin{flalign}
C_{v,u}&=B_1\log_2\Big(1+\frac{P_v}{\sigma_{vm}^2}|\mathbf{h}_{vm}|^2\Big), \label{mainca} \\ 
C_{e,u}&=B_1\log_2\det\bigg(\mathbf{I}_{N_e}+\frac{P_v\mathbf{h}_{ve}\mathbf{h}_{ve}^{\dagger}}{\sigma_{ve}^2\mathbf{I}_{N_e}+\frac{P_m}{N_m-1}\mathbf{H}_{1}\mathbf{H}_{1}^{\dagger}}\bigg), \label{wiretapca}
\end{flalign}
\end{subequations}
where $B_1$ is the uplink bandwidth and $\mathbf{H}_1=\mathbf{H}_{me}\mathbf{G}_u\in\mathbb{C}^{N_e\times (N_m-1)}$. $C_{v,u}$ is the uplink channel capacity between the ICV and the SBS that can be achieved via $\mathbf{w}_m$, and $C_{e,u}$ is the uplink capacity between the ICV and Eve. Note that $P_v$ and $P_m$ are normalized by the bandwidth.

The equality of Eqn. (\ref{lbscu}) holds if and only if $\{C_{v,u}-C_{e,u}\}$ is always nonnegative in all channel states. However, due to the lack of $\mathbf{H}_{me}$, we cannot determine whether an instantaneous secrecy rate is nonnegative or not, and thus we resort to derive a lower bound of the real ergodic secrecy rate as an ergodic secrecy rate. According to \cite{Liu2019ArtificialNoisyMIMO}, we know $C_{v,u}$ is larger than $C_{e,u}$ with a high probability because of AN signals. The uplink ergodic secrecy rate can be written as
\begin{flalign}\label{luesc}
\bar{R}_{u}=\text{E}_{\mathbf{h}_{vm}}[C_{v,u}]-\text{E}_{\mathbf{h}_{vm},\mathbf{h}_{ve},\mathbf{H}_{me}}[C_{e,u}]. 
\end{flalign}
Then, we will derive a theoretical expression of the ergodic secrecy rate $\bar{R}_u$ for task partition, i.e., use $\bar{R}_u$ in P1 instead of $R_u$, which is deduced in the following proposition.

\vspace{0.1in}
\begin{proposition}
The ergodic secrecy rate of $R_u$, i.e., $\bar{R}_u$ is 
\begin{flalign}\label{ESRu}
\bar{R}_u  = &   B_1 \big \{\Phi(\rho_1)+ C(\mathbf{H}_1,\rho_2) - \Psi(\mathbf{H}_2,\rho_2,P_v)\big\},
\end{flalign}
where 
\begin{flalign}
&\Phi(\rho)=\frac{1}{\ln(2)}\exp(\rho) \sum_{k=0}^{N_m-1}E_{k+1}(\rho), \label{mainec}\\ 
&\rho_1=\frac{\sigma_{vm}^2}{P_v}, \quad \rho_2 =\frac{P_m}{\sigma_{ve}^2(N_m-1)},
\end{flalign}
$\mathbf{H}_1=\mathbf{H}_{me}\mathbf{G}_u\in\mathbb{C}^{N_e\times (N_m-1)}$, $\mathbf{H}_2=[\mathbf{h}_{ve}, \mathbf{H}_{1}]\in\mathbb{C}^{N_e\times N_m}$,
\begin{flalign}\label{erCm1}
&C(\mathbf{A},\rho) \notag \\
&=\frac{\exp(1/\rho)}{\ln(2)}\sum_{k=0}^{n-1}\sum_{l=0}^{k}\sum_{i=0}^{2l}\bigg\{\frac{(-1)^i(2l)!(m-n+i)!}{2^{2k-i}l!i!(m-n+l)!}\\ \notag
&\times \binom{2k-2l}{k-l}\binom{2l+2m-2n}{2l-i}\sum_{j=0}^{m-n+i}E_{j+1}(1/\rho)\bigg\},
\end{flalign}
$n=\min(a,b)$, $m=\max(a,b)$ for any $\mathbf{A}\in \mathbb{C}^{a\times b}$, $E_{\tau}(z)$ is the exponential integral of order $\tau$ defined by
\begin{flalign}\label{EN}
E_{\tau}(z)=\int_1^{+\infty}e^{-zx}x^{-\tau}dx,~~~\tau=0,1,...,\text{Re}(z)>0,
\end{flalign}
and $\text{Re}(z)$ is the real part of $z$. Finally, we have
\begin{flalign}\label{esc1}
\Psi(\mathbf{H}_2,\rho_2,P_v)=K\sum_{k=1}^{N_e}\det\{\mathbf{\Theta}(k,\mathbf{Q})\},
\end{flalign}
where we have $N_m\times N_m$ matrix $\mathbf{Q}=\text{diag}(P_v/\sigma_{ve}^2,\rho_2,...,\rho_2)$. The $i$-th row and $j$-th column element of the $N_m \times N_m$ real matrix $\mathbf{\Theta}(k,\mathbf{Q})$ are defined as
\begin{flalign}\label{Matrix2}
&\big[\mathbf{\Theta}(k,\mathbf{Q})\big]_{i,j} \notag \\
&=\begin{cases}
(-1)^{d_i}(j-1+d_i)!/\mu_{(c_i)}^{j+d_i},&\text{$j=1,\dots,N_e$, $j \neq k$},\\
\frac{(-1)^{d_i}}{\ln(2)}e^{\mu_{(c_i)}}(j-1+d_i)!\Omega,&\text{$j=1,\dots,N_e$, $j=k$},\\
[N_m-j]_{d_i}\{\mu_{(c_i)}^{N_m-d_i-j}\},&\text{$j=N_e+1,\dots,N_m$},
\end{cases}
\end{flalign}
where $[y]_x=y(y-1)...(y-x+1)$, $[y]_0=1$,
\begin{flalign}
& \Omega=\sum_{t=0}^{j-1+d_i}\frac{\Gamma(t-j+1-d_i,\mu_{(c_i)})}{\mu_{(c_i)}^{t+1}},\\
&\Gamma(a,x)=\int_x^{+\infty}\exp(-z)z^{a-1}dz, \\ 
&K=\frac{(-1)^{N_e(N_m-N_e)}\prod_{i=1}^2\mu_i^{\nu_iN_e}}{\Gamma_{N_e}(N_e)\prod_{i=1}^{2}\Gamma_{\nu_i}(\nu_i)(\mu_1-\mu_2)^{\nu_1\nu_2}},
\end{flalign}
$\Gamma_{\alpha}(\beta)=\prod_{i=1}^\alpha(\beta-i)!$, and $\mu_1>\mu_2$ are the distinct eigenvalues of $\mathbf{Q}^{-1}$, whose numbers are $\nu_1$ and $\nu_2$, respectively, so that $\sum_{i=1}^2\nu_i=N_m$. Let $c_i$ denote a unique integer such that 
\begin{flalign}
\nu_1+...+\nu_{c_i-1}<i\leq \nu_1+...+\nu_{c_i},
\end{flalign}
and 
\begin{flalign}
d_i=\sum_{j=1}^{c_i}\nu_j-i.
\end{flalign}
Note that when $\mu_1=\mu_2$, i.e., $\rho_2=P_v/\sigma_{ve}^2$, we have $\Psi(\mathbf{H}_2,\rho_2,P_v)=C(\mathbf{H}_2,\rho_2)$.
\end{proposition}

\begin{IEEEproof}
See in Appendix A.  
\end{IEEEproof}

\vspace{0.1in}
Although Eqn. (\ref{ESRu}) has no integral expression except for several special functions, it is a complex equation, which yields a lot of computation overhead. We provide a closed-form expression of the lower bound of $\bar{R}_u$ in a high SNR region, i.e., when $P_m/\sigma_{ve}^2$ is high, as given in the following corollary. The lower bound is a common metric in security as it represents the worst case scenario.

\begin{corollary}
In a high SNR region, i.e., if $P_m/\sigma_{ve}^2$ is high, the lower bound of $\bar{R}_u$, i.e., $\tilde{R}_u$ is
\begin{flalign}\label{lbescu}
\tilde{R}_u =B_1\Big\{\Phi(\rho_1)+\sum_{i=0}^{N_e-1}\psi(N_m-1-i)-\log_2(\chi_1)\Big\},
\end{flalign}
where 
\begin{flalign}
&\psi(x)=\frac{1}{\ln(2)}\Big(-\xi+\sum_{r=1}^{x-1}\frac{1}{r}\Big), \label{psi} \\
& \chi_1=A_1+ \frac{P_v}{\sigma_{ve}^2\rho_2}\times A_2,\\
& A_1 =\sum_{i=1}^{N_e} \sum_{j=2}^{N_m}[N_m-i+1]_i\binom{N_m-j}{i-1}+1,\\
&A_2=\sum_{i=1}^{N_e}[N_m-i+1]_i\binom{N_m-1}{i-1},
\end{flalign}
$\xi=0.577215...$ is the Euler's constant, $[y]_x=y(y-1)...(y-x+1)$, $[y]_0=1$, $\Phi(\rho)$ is defined in Eqn. (\ref{mainec}), and $\binom{x}{y}=\frac{x!}{(x-y)!y!}$. Note that when $N_m-j<i-1$, we set $\binom{N_m-j}{i-1}=0$ in $A_1$.
\end{corollary}

\begin{IEEEproof}
See in Appendix B.  
\end{IEEEproof}

\begin{remark}[Uplink power gain of SBS]
$P_m$ only affects $\log_2(\chi_1)$ in Eqn. (\ref{lbescu}). $\chi_1$ will decrease and approach to $A_1$ with an increasing $P_m$. Thus, $\tilde{R}_u$ grows with an increasing $P_m$, and then approaches to a constant that equals to Eqn. (\ref{lbescu}) after replacing $\chi_1$ with $A_1$.
\end{remark}

\subsection{Downlink Ergodic Secrecy Rate $\bar{R}_d$}\label{downprecoding}
For downlink security purpose, the SBS uses a joint MRT and nullspace-based AN technology. More specifically, the MRT vector is $\mathbf{w}_d=\mathbf{h}_{mv}^{\dagger}/|\mathbf{h}_{mv}|$, and the received signals at the ICV and Eve are formulated as
\begin{subequations}
\begin{flalign}
y_{mv}&=\mathbf{h}_{mv}\mathbf{w}_dx_d+\mathbf{h}_{mv}\mathbf{G}_d\mathbf{z}_d+n_{mv} \notag \\
 & =\mathbf{h}_{mv}\mathbf{w}_dx_d+n_{mv}, \\
\mathbf{y}_{me}&=\mathbf{H}_{me}\mathbf{w}_dx_d+\mathbf{H}_{me}\mathbf{G}_d\mathbf{z}_d+\mathbf{n}_{me},
\end{flalign}
\end{subequations}
where $\mathbf{G}_d$ is an $N_m\times(N_m-1)$ matrix that lies in the nullspace of $\mathbf{h}_{mv}$, $\mathbf{z}_d$ is a complex Gaussian AN signal obeying $\mathcal{CN}(\mathbf{0},\frac{P_m}{N_m}\mathbf{I}_{N_m-1})$, and $\text{E}[x_dx_d^{\dagger}]=P_m/N_m$. $n_{mv}$ is an AWGN variable obeying $\mathcal{CN}(0,\sigma_{mv}^2)$, and $\mathbf{n}_{me}$ is an AWGN vector obeying $\mathcal{CN}(\mathbf{0},\sigma_{me}^2\mathbf{I}_{N_e})$. Similar to the uplink phases, we will use ergodic secrecy rate for task partition, i.e., to use $\bar{R}_d$ in P1 instead of $R_d$. The real ergodic secrecy rate in downlink channel is 
\begin{flalign}\label{escd}
\ddot{R}_{d}=\text{E}_{\mathbf{h}_{mv},\mathbf{H}_{me}}[C_{v,d}-C_{e,d}]^{+},
\end{flalign}
where
\begin{subequations}
\begin{flalign}
C_{v,d}&=B_2\log_2\Big(1+\frac{P_m}{N_m\sigma_{mv}^2}|\mathbf{h}_{mv}|^2\Big), \label{dmc} \\ 
C_{e,d}&=B_2\log_2\det\bigg(\mathbf{I}_{N_e}+\frac{P_m\mathbf{h}_{1}\mathbf{h}_{1}^{\dagger}}{N_m\sigma_{me}^2\mathbf{I}_{N_e}+P_m\mathbf{H}_{3}\mathbf{H}_{3}^{\dagger}}\bigg), \label{dwc}
\end{flalign}
\end{subequations}
$B_2$ is the downlink bandwidth, $\mathbf{h}_1=\mathbf{H}_{me}\mathbf{w}_{d}\in\mathbb{C}^{N_e\times 1}$, and $\mathbf{H}_3=\mathbf{H}_{me}\mathbf{G}_d\in\mathbb{C}^{N_e\times (N_m-1)}$. 

Similar to the uplink phases, the downlink ergodic secrecy rate is formulated as
\begin{flalign}\label{lbesc}
\bar{R}_{d}=\text{E}_{\mathbf{h}_{mv}}[C_{v,d}]-\text{E}_{\mathbf{h}_{mv},\mathbf{H}_{me}}[C_{e,d}]. 
\end{flalign}

We provide an exact expression of $\bar{R}_{d}$ to measure the downlink ergodic secrecy rate in the following proposition.

\vspace{0.1in}
\begin{proposition}
The ergodic secrecy rate of $R_d$, i.e., $\bar{R}_d$ is 
\begin{flalign}\label{ESR1}
\bar{R}_d  = & B_2 \{\Phi(\rho_3)+ C(\mathbf{H}_3,\rho_4) -  C(\mathbf{H}_4,\rho_4)\},
\end{flalign}
where 
\begin{flalign}\label{ESR1}
&\rho_3=\frac{\sigma_{mv}^2N_m}{P_m}, \quad \rho_4 =\frac{P_m}{\sigma_{me}^2N_m},
\end{flalign}
$\mathbf{H}_4=[\mathbf{h}_{1}, \mathbf{H}_{3}]\in\mathbb{C}^{N_e\times N_m}$, and $C(\mathbf{A},\rho)$ is defined in Eqn. (\ref{erCm1}).
\end{proposition}

\begin{IEEEproof}
The proof in \cite[Th. 3] {Liu2017SecrecyCapacityAnalysis} shows a general case, where the numbers of antennas in receivers and eavesdroppers are arbitrary. Proposition 2 is a special case that the number of antennas at receivers is one, which can be deduced from \cite[Th. 3] {Liu2017SecrecyCapacityAnalysis} via changing the variable of the number of antennas. 
\end{IEEEproof}

Also, we provide a closed-form expression of the lower bound of $\bar{R}_d$ in a high SNR region, i.e., when $P_m/\sigma_{me}^2$ is high, as in the following corollary.

\begin{corollary}
In a high SNR region, i.e., when $P_m/\sigma_{me}^2$ is high, the lower bound of $\bar{R}_d$, i.e., $\tilde{R}_d$ is
\begin{flalign}\label{lbdec}
\tilde{R}_d= B_2 \{\Phi(\rho_3)+\sum_{i=0}^{N_e-1}\psi(N_m-1-i)-\log_2(\chi_2)\},
\end{flalign}
where $\chi_2=N_m!/(N_m-N_e)!$, $\psi(x)$ is defined in Eqn. (\ref{psi}), and $\Phi(\rho)$ is defined in Eqn. (\ref{mainec}).
\end{corollary}

\begin{IEEEproof}
See in Appendix C. 
\end{IEEEproof}

\begin{remark}[Downlink power gain of SBS]
$P_m$ only affects $\Phi(\rho_3)$ in Eqn. (\ref{lbdec}). $\Phi(\rho_3)$ increases monotonically with an increasing $P_m$, so does $\tilde{R}_d$, meaning that the downlink power gain of $P_m$ is larger than that of uplink.
\end{remark}

\subsection{Computation Task Partition with $\bar{R}_u$ and $\bar{R}_d$}

The ICV will use $\{\bar{R}_u, \bar{R}_d, a_v,a_m\}$ for the computation task partition. First, we introduce an auxiliary function to represent estimated latency of the MEC server that processes $(1-\eta)M$ bits data as
\begin{flalign}
T_{\text{MEC}}(\eta)=T_m(\eta)+T_u(\eta)+T_d(\eta),
\end{flalign}
where $T_m(\eta)=\frac{(1-\eta) M}{a_m}$, $T_u(\eta)=\frac{\beta(1-\eta)M}{\bar{R}_u}$, and $T_d(\eta)=\frac{\alpha(1-\eta)M}{\bar{R}_d}$. As the descriptions given in the problem formulation, $\alpha$ accounts for a ratio of output to input bits offloaded to the MEC server, and $\beta$ is a compression ratio of the uploaded data. Hence, with the auxiliary function and parameters $\{\bar{R}_u, \bar{R}_d, a_v,a_m\}$, P1 can be formulated as
\begin{flalign}\label{GSNR}
\text{P2: } &\min_{0 \leq \eta \leq 1}\max\{T_v(\eta),T_{\text{MEC}}(\eta)\},
\end{flalign}
where $T_v(\eta)=\eta M/a_v$ is the computing time of the ICV. The optimal $\eta$ has a closed-form solution, as seen in the following proposition.

\begin{proposition}
The optimal computation task partition ratio $\eta^*$ can be expressed as 
\begin{equation}\label{oall}
\eta^*=1-\frac{1}{a_va_1+1},
\end{equation}
where
\begin{flalign}\label{eta2}
a_1=\frac{1}{a_m}+\frac{\beta}{\bar{R}_u}+\frac{\alpha}{\bar{R}_d}.
\end{flalign}
\end{proposition}

\begin{IEEEproof}
See in Appendix D.
\end{IEEEproof}

\begin{remark}[Advantage of edge computing]
Since $a_v>0$ and $a_1>0$, it is obvious that $0<\eta^*<1$ highlights the advantage of edge computing, i.e., the computation strategy of an ICV should offload a part of computation tasks to an SBS, while computing the rest of computation tasks in an on-board computer in parallel.
\end{remark}

\begin{remark}[Cask principle]
When $a_v\rightarrow 0$, we have $\eta^*\rightarrow 0$, which means that if the computing capacity of an ICV is very small, the total computation task will be uploaded to the SBS. While $a_1\rightarrow +\infty$, we have $\eta^*\rightarrow 1$, which means that the total computation task will be processed at the local ICV computer, and the ability of edge computing follows the ``cask principle'', where the smallest one of the metrics decides the capability of edge computing. That is, if any of $a_m\rightarrow 0$, $\bar{R}_u\rightarrow 0$, or $\bar{R}_d\rightarrow 0$ is achieved, we have $a_1\rightarrow +\infty$ and total computation task will be processed at the local ICV computer.
\end{remark}

\section{UPLINK/DOWNLINK ADAPTIVE WIRETAP CODING}\label{coding}
In the above section, we used ergodic secrecy rates of uplink and downlink for the task partition, because the ergodic secrecy rates can be calculated without instantaneous CSIs of the main and wiretap channels. The ergodic secrecy rate is just a global secrecy metric that can not be used for wiretap coding. In uplink and downlink transmissions, the instantaneous CSIs of the main channels, i.e., $\mathbf{h}_{vm}$, $\mathbf{h}_{mv}$, and $\mathbf{H}_{mm}$ can be obtained via channel estimation, while the instantaneous CSIs of wiretap channels are still unavailable since Eve is silent. With instantaneous CSIs of the main channels and statistical CSIs of wiretap channels, a common way in PHY-layer security is to consider the secrecy outage probability with wiretap coding \cite{Yang2018SmallCellAssisted} \cite{Zhou2011RethinkingSecrecyOutage}, i.e., to adaptively adjust the wiretap code rates for both uplink and downlink transmissions with the given secrecy outage probability constraints.  

\subsection{Uplink Adaptive Wiretap Code Rate}\label{upcoding}
The ICV obtains $\mathbf{h}_{vm}$ and the SBS uses the MRC and AN schemes as described in Section \ref{upprecoding}. The instantaneous uplink secrecy rate can be written as 
\begin{flalign}\label{scup}
R_{1}=[C_{v,u}-C_{e,u}]^{+},
\end{flalign}
where $C_{v,u}$ and $C_{e,u}$ are formulated in Eqns. (\ref{mainca}) and (\ref{wiretapca}). Due to the fact that ICV cannot access Eve's CSI, it adopts secrecy outage probability as a performance metric for adaptive wiretap coding, which is defined as the probability that the target wiretap code rate of secure transmissions, i.e., $R_u$ is larger than the secrecy rate $R_1$. From \cite[Eq. (4)]{Zhou2011RethinkingSecrecyOutage}, we know that the secrecy outage probability is expressed as 
\begin{flalign}\label{uso}
P_{\text{out}}(R_u)&=P(R_1\leq R_u \big| \text{message transmission}) \notag \\
&=P(C_{e,u}\geq C_{v,u}- R_u\big),
\end{flalign}
which means that the secrecy outage probability is the conditional probability based on the reliability of transmitted codewords, i.e., SBS is able to decode correctly with the rate of transmitted codewords, where the rate can be up to $C_{v,u}$. Based on Wyner's coset based encoding theory, to achieve physical layer security, the encoder should choose two rates, namely, the rate of transmitted codewords of common messages ($C_{v,u}$), and the rate of confidential information, namely, wiretap code rate ($R_u$)\cite{Oggier2016}. Since we assume that ICV has perfect knowledge about the instantaneous CSI of $\mathbf{h}_{vm}$ within coherence time, it is possible to use an adaptive rate of transmitted codewords that equals to $C_{v,u}$. The remaining work is to find an appropriate $R_u$.

Based on the secrecy outage probability, we present an effective secrecy rate as a secrecy metric, which means an average rate secretly received at SBS over many transmission bursts with a wiretap code rate $R_u$, which is expressed as
\begin{flalign}\label{effsr}
\hat{R}_u(R_u)=\{1-P_{\text{out}}(R_u)\}R_u.
\end{flalign}
We can emulate Eqn. (\ref{effsr}) via Monte Carlo simulations over all realizations of $\mathbf{H}_{me}$. However, in order to adaptively adjust wiretap code rate $R_u$ to maximize $\hat{R}_u$, it is necessary to deduce an effective secrecy rate $\hat{R}_u$. The expression is presented in Proposition 4.

\begin{proposition}
The effective secrecy rate of $R_u$, i.e., $\hat{R}_u$, can be expressed as
\begin{flalign}\label{exeffsr}
\hat{R}_u(R_u)=\{1-F_{Z_1}(\phi_1)\}R_u,
\end{flalign}
where $\phi_1=\frac{P_m}{P_v(N_m-1)} (2^{C_{v,u}-R_u}-1)$,
\begin{flalign}\label{tout}
&F_{Z_1}(z)=\exp(-a_1 z)\sum_{k=0}^{N_e-1}\frac{A_k(z)}{k!}(a_1 z)^k, \notag \\
& A_k(z)=\frac{\sum_{n=0}^{N_e-k-1}\binom{N_m-1}{n}z^n}{(1+z)^{N_m-1}},
\end{flalign}
and $Z_1=\mathbf{h}_{ve}^{\dagger}(a_1\mathbf{I}_{N_e}+\mathbf{H}_{1}\mathbf{H}_{1}^{\dagger})^{-1}\mathbf{h}_{ve}$ represents a random variable, where $a_1=\frac{\sigma_{ve}^2(N_m-1)}{P_m}$. It is obvious that $F_{Z_1}(\phi_1)$ is the expression of secrecy outage probability of $R_u$.
\end{proposition}

\begin{IEEEproof}
See in Appendix E.
\end{IEEEproof}

\vspace{0.1in}
Proposition 4 provides exact expressions of Eqn. (\ref{effsr}) and $P_{\text{out}}(R_u)=F_{Z_1}(\phi_1)$. We can further simplify the expression if Eve has a very high SNR, i.e., $P_m/\sigma_{ve}^2\rightarrow \infty$, as shown in Corollary 3.

\begin{corollary}
When $P_m/\sigma_{ve}^2\rightarrow \infty$, the effective secrecy rate $\hat{R}_u$ can be expressed approximately as
\begin{flalign}\label{outageapp}
\hat{R}_u(R_u)\simeq\{1-Q_{Z_1}(\phi_1)\}R_u,
\end{flalign}
where $\phi_1=\frac{P_m}{P_v(N_m-1)} (2^{C_{v,u}-R_u}-1)$,
\begin{flalign}\label{ISC2}
Q_{Z_1}(z)=\frac{\sum_{n=0}^{N_e-1}\binom{N_m-1}{n}z^n}{(1+z)^{N_m-1}},
\end{flalign}
and $Z_1=\mathbf{h}_{ve}^{\dagger}(\mathbf{H}_{1}\mathbf{H}_{1}^{\dagger})^{-1}\mathbf{h}_{ve}$ represents a random variable.
\end{corollary}

\begin{IEEEproof}
When $P_m/\sigma_{ve}^2\rightarrow \infty$, we get $a_1=0$. Then, $Q_{Z_1}(z)=A_0(z)$, and thus it is easy to obtain Eqn. (\ref{outageapp}). An approximated expression of secrecy outage probability, i.e., Eqn. (\ref{ISC2}), is also given in \cite[Eq. (45)] {Ng2011}.
\end{IEEEproof}

\vspace{0.1in}
\subsubsection{Effective secrecy rate maximization} 
Based on Proposition 4, we propose an adaptive wiretap coding scheme to maximize effective secrecy rate as follows.
\begin{flalign}
(\hat{R}_u^*,R_u)=\max_{P_{\text{out}}(R_{u})\leq \varepsilon_u}\hat{R}_u(R_u),
\end{flalign}
where the optimal $\hat{R}_u$, i.e., $\hat{R}_u^*$, and the corresponding $R_u$ can be obtained via one-dimensional search on the function $\hat{R}_u(R_u)$ with a secrecy outage probability constraint $\varepsilon_u$. 

\subsubsection{Secrecy outage limitation} 
An alternative adaptive wiretap coding scheme considers only secrecy outage probability. With a secrecy outage limitation $P_{\text{out}}(R_{u})\leq \varepsilon_u$, we can maximize $R_{u}$ via the inverse operation of $\varepsilon_u=P_{\text{out}}(R_{u})$ due to the fact that $P_{\text{out}}(R_u)=F_{Z_1}(\phi_1)$ is a decreasing function of $R_u$. The method is much simpler than that of effective secrecy rate maximization, which does not require any search algorithms.

\begin{figure*}[t]
\centering
\subfigure[Uplink ergodic secrecy rates. ]{
\label{ersru} 
\includegraphics[width=0.45\textwidth]{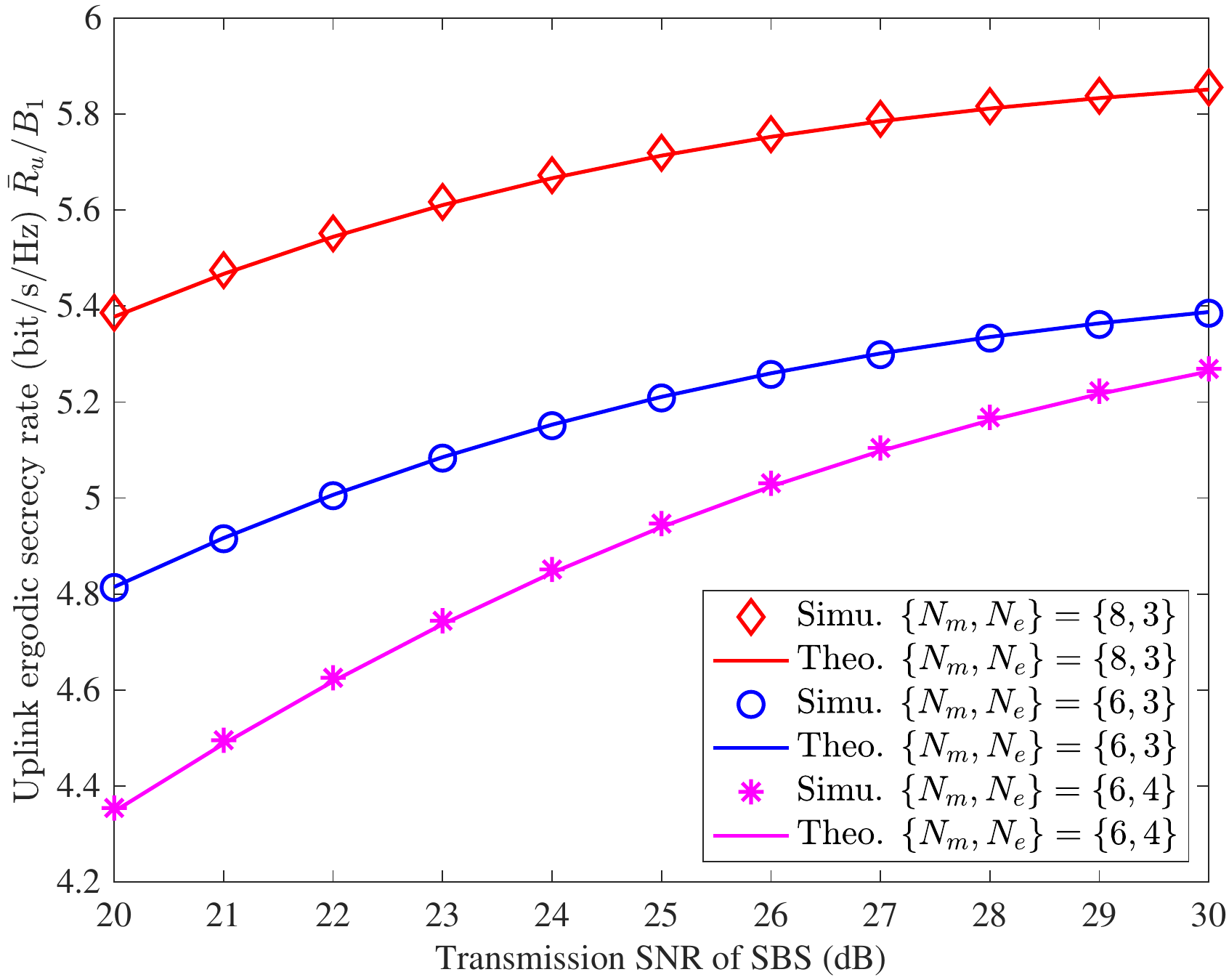}}
\hspace{0.7cm}
\subfigure[Downlink ergodic secrecy rates.]{
\label{ersrd} 
\includegraphics[width=0.45\textwidth]{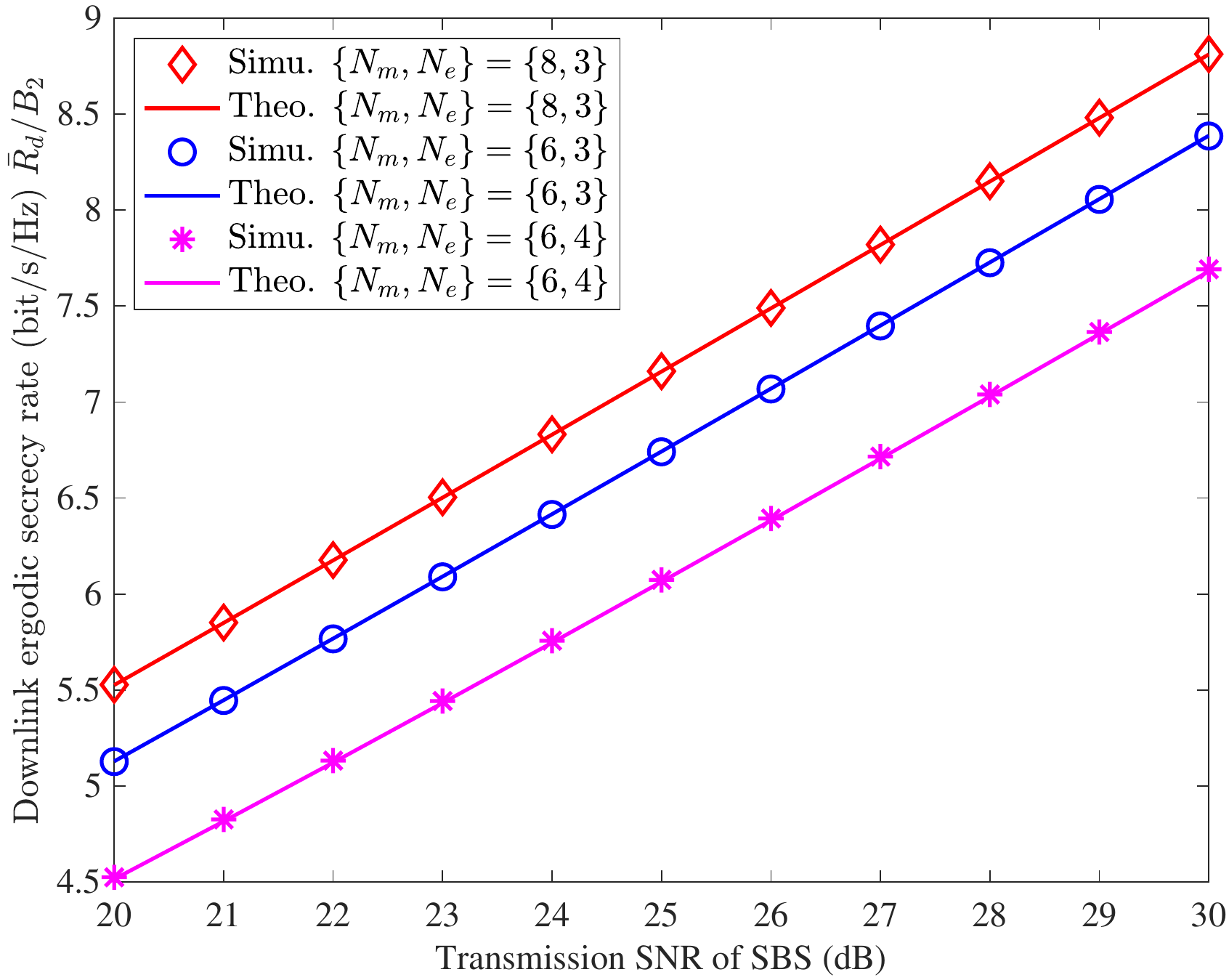}}
\caption{Numerical results of uplink and downlink ergodic secrecy rates in terms of SBS transmission SNR, where ICV transmission SNR is 10 dB in uplink.}
\label{ers} 
\end{figure*}

\subsection{Downlink Adaptive Wiretap Code Rate}\label{docoding}
The SBS obtains $\mathbf{h}_{mv}$ and uses the MRT and AN schemes as described in Section \ref{downprecoding}. In this case, the instantaneous downlink secrecy rate can be formulated as 
\begin{flalign}\label{scup}
R_{2}=[C_{v,d}-C_{e,d}]^{+},
\end{flalign}
where $C_{v,d}$ and $C_{e,d}$ are formulated in Eqns. (\ref{dmc}) and (\ref{dwc}). Similar to Proposition 4, effective secrecy rate $\hat{R}_d$ can be expressed as 
\begin{flalign}\label{outaged}
\hat{R}_d(R_d)=\{1-P_{\text{out}}(R_d)\}R_d.
\end{flalign}

\begin{proposition}
The effective secrecy rate of $R_d$, i.e., $\hat{R}_d$, can be expressed as
\begin{flalign}\label{outage}
\hat{R}_u(R_u,C_{v,u})=\{1-F_{Z_2}(\phi_2)\}R_u,
\end{flalign}
where $\phi_2=2^{C_{v,d}-R_d}-1$,
\begin{flalign}
&P_{Z_2}(z)=\exp(-za_2)\sum_{k=0}^{N_e-1}\frac{A_k(z)}{k!}(za_2)^k, \notag \\
&A_k(z)=\frac{\sum_{n=0}^{N_e-k-1}\binom{N_m-1}{n}z^n}{(1+z)^{N_m-1}},
\end{flalign}
$Z_2=\mathbf{h}_{1}^{\dagger}(a_2\mathbf{I}_{N_e}+\mathbf{H}_{3}\mathbf{H}_{3}^{\dagger})^{-1}\mathbf{h}_{1}$ represents a random variable, and $a_2=\frac{N_m\sigma_{me}^2}{P_m}$. The adaptive channel coding with $C_{v,d}$ is used in SBS with the knowledge of $\mathbf{h}_{mv}$. 
\end{proposition}

\begin{IEEEproof}
Similar to Proposition 4.
\end{IEEEproof}

\vspace{0.1in}
With effective secrecy rate $\hat{R}_d$ and secrecy outage probability $F_{Z_2}(\phi)$, an SBS can execute effective secrecy rate maximization or adaptive rate adjustment with the secrecy outage limitations, similar to that in the uplink phase. 

\subsection{Computation Complexity Analysis}

Next, let us discuss about the computation complexity of the proposed scheme as follows.

The precoding/coding process of PHY-layer security includes uplink/downlink precoding and adaptive wiretap coding. The computation complexity of the wiretap codebook encoding and decoding is $O(M)$, where $M$ is the number of lattice points \cite{Oggier2016} \cite{Sloane1981}. As shown in Propositions 4 and 5, adaptive adjustment of wiretap coding rates can be achieved by Golden-section searching algorithm with its computation complexity $O\big(\log (1/\epsilon)\big)$, where $\epsilon$ is the required accuracy. For uplink precoding, SBS should generate the MRC receiver $\mathbf{w}_m=\mathbf{h}_{vm}^{\dagger}/|\mathbf{h}_{vm}|$ and the nullspace $\mathbf{G}_u$, which need $O(4N_m)$ and $O(3N_m^2+4N_m)$ time overhead, respectively. Similarly, for downlink precoding, SBS needs $O(4N_m)$ and $O(2N_m^2+3N_m)$ time overhead to generate the MRT vector $\mathbf{w}_d=\mathbf{h}_{mv}^{\dagger}/|\mathbf{h}_{mv}|$ and the nullspace $\mathbf{G}_d$, respectively \cite{Raz2002}. In conclusion, the computation complexity of the precoding algorithms is polynomial and thus practical in 5G communication systems.

Note that computation task partition ratio is calculated by ergodic secrecy rates $\bar{R}_u$ and $\bar{R}_d$, which can be viewed as the given system parameters because $\bar{R}_u$ and $\bar{R}_d$ are determined by the pre-defined parameters $\{a_v,a_m,N_m,N_e,P_m,P_v\}$ as shown in Propositions 1 and 2, with no dependence on the instantaneous CSIs of both main and wiretap channels. Also, the optimal computation task partition ratio between onboard computers and MEC servers has been formulated in a closed-form solution, as shown in Proposition 3, which can be calculated with very little computation overhead. 

\begin{figure*}[t]
\centering
\subfigure[Uplink effective secrecy rates. ]{
\label{efsru} 
\includegraphics[width=0.45\textwidth]{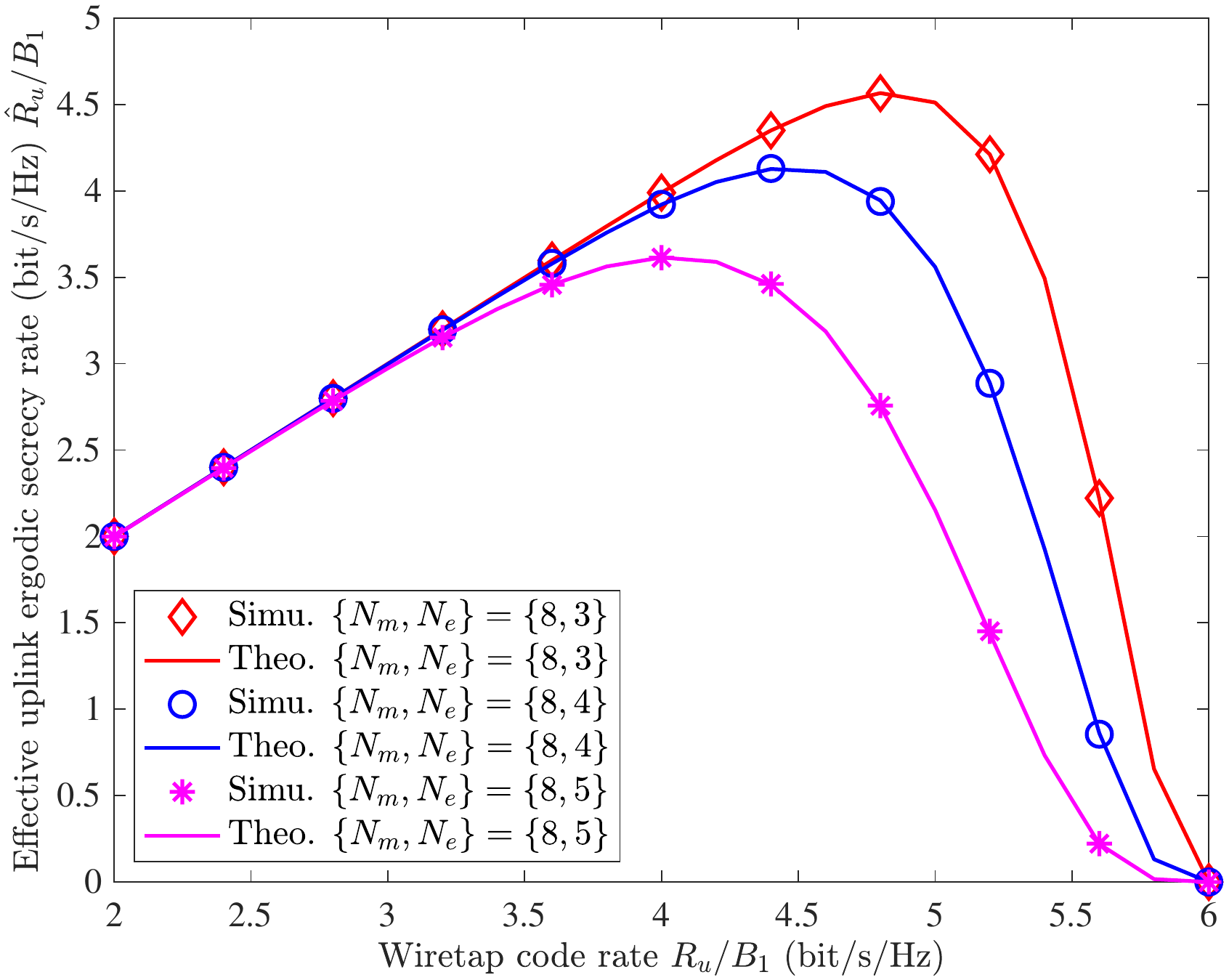}}
\hspace{0.7cm}
\subfigure[Downlink effective secrecy rates.]{
\label{efsrd} 
\includegraphics[width=0.45\textwidth]{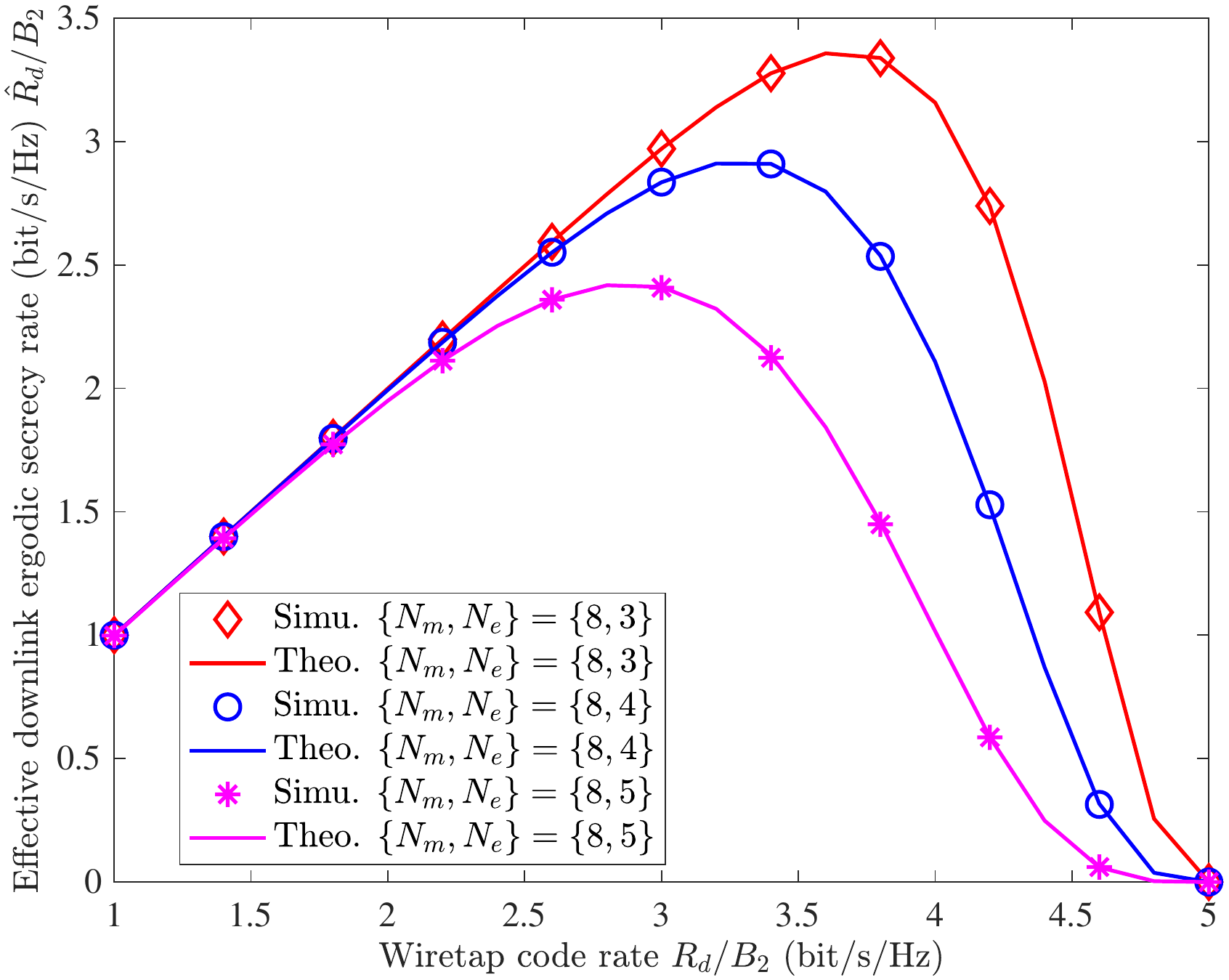}}
\caption{Numerical results of uplink and downlink effective secrecy rates in terms of wiretap code rates, where transmission SNRs of ICV and SBS are 10 dB and 20 dB, $C_{v,u}/B_1=6$ bit/s/Hz, and $C_{v,d}/B_2=5$ bit/s/Hz.}
\label{efsr} 
\end{figure*}

\section{NUMERICAL AND SIMULATION RESULTS}\label{simulation}

In this section, we first examine Propositions 1, 2, 4, and 5 in Figs. \ref{ersru}, \ref{ersrd}, \ref{efsru}, and \ref{efsrd}, respectively. These figures show the good agreements between theoretical results (Theo.) and Monte Carlo simulation results (Simu.) from $10^5$ independent runs. 

In particular, Fig. \ref{ersru} illustrates the impact of transmission SNR of SBSs on uplink ergodic secrecy rates, where theoretical results were calculated from Proposition 1, and Monte Carlo simulations were done based on Eqn. (\ref{luesc}). We can see that ergodic secrecy rates increase with an increasing transmission SNR of SBSs, and will reach to a constant, which is consistent with the discussions in Remark 1. Also, more antennas at SBSs can yield a better performance, while an increasing number of Eve's antennas will reduce ergodic secrecy rates. Fig. \ref{ersrd} shows the impact of SNR on the downlink, where theoretical results were calculated from Proposition 2, and Monte Carlo simulations were based on Eqn. (\ref{lbesc}). We can see that ergodic secrecy rates grow almost linearly with SNR, which is also consistent with Remark 2. The ergodic secrecy rate is a representative performance index of the proposed scheme. As shown in Figs. \ref{ersru} and \ref{ersrd}, uplink and downlink ergodic secrecy rates are approximately 6 bit/s/Hz and 9 bit/s/Hz, respectively, i.e., 120 Mbps and 180 Mbps for secrecy transmission over a 20 MHz channel.

\begin{figure*}[h]
\begin{minipage}[t]{0.47\linewidth}
\centering
\includegraphics[width=1\linewidth]{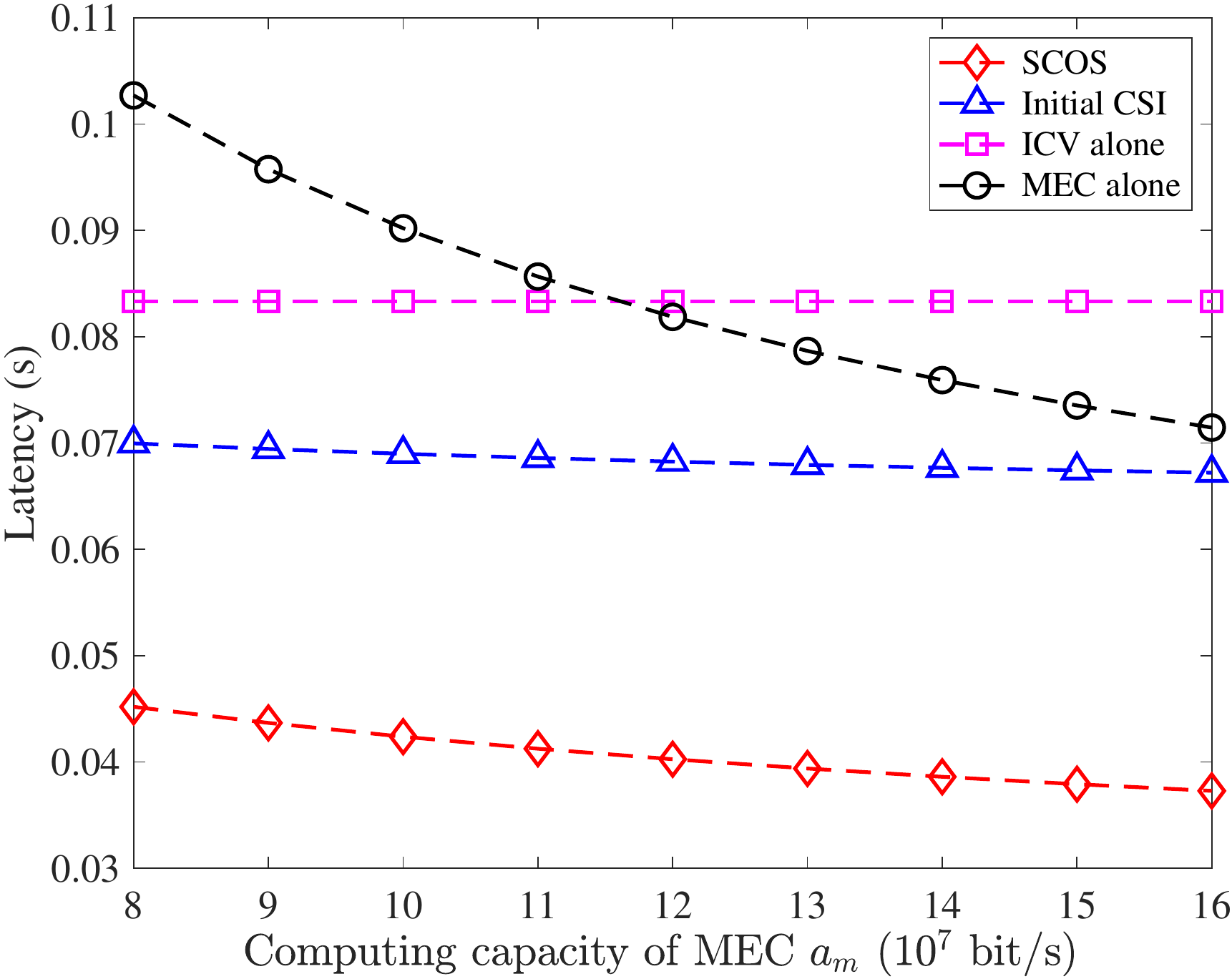}
\caption{Latency in terms of MEC server computing capacity, i.e., $a_m$, where we set $a_v=6\times 10^7$ bit/s, $N_m=10$, $N_e=4$, and transmission SNRs of SBS and ICV are 20 dB and 10 dB, respectively.}\label{sim1}
\end{minipage}
\hspace{0.2cm}
\begin{minipage}[t]{0.47\linewidth}
\centering
\includegraphics[width=1\linewidth]{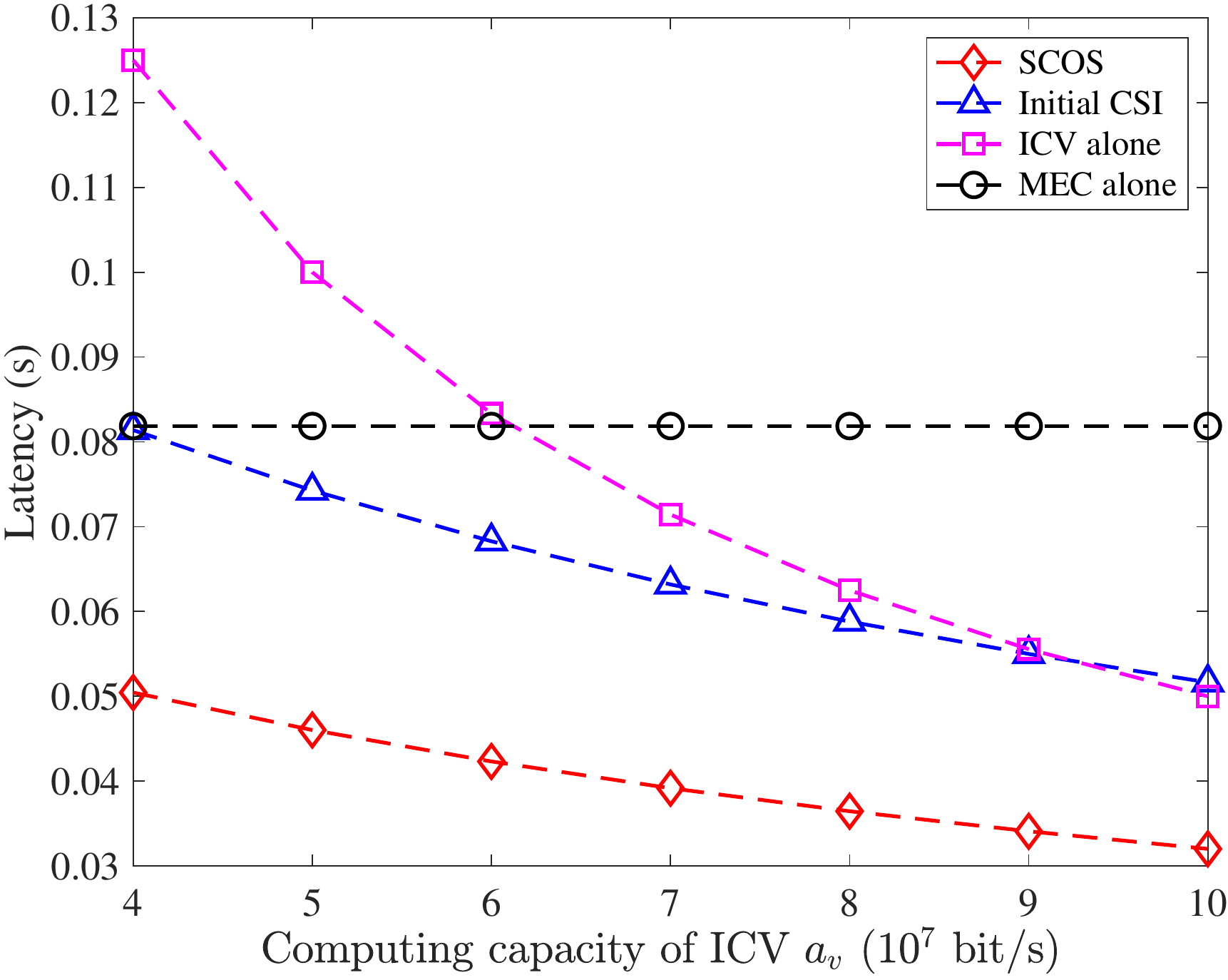}
\caption{Latency in terms of ICV computing capacity, i.e., $a_v$, where we set $a_m=12 \times 10^7$ bit/s, $N_m=10$, $N_e=4$, and transmission SNRs of SBS and ICV are 20 dB and 10 dB, respectively.}\label{sim2}
\end{minipage}
\end{figure*}

\begin{figure*}[h]
\begin{minipage}[t]{0.47\linewidth}
\centering
\includegraphics[width=1\linewidth]{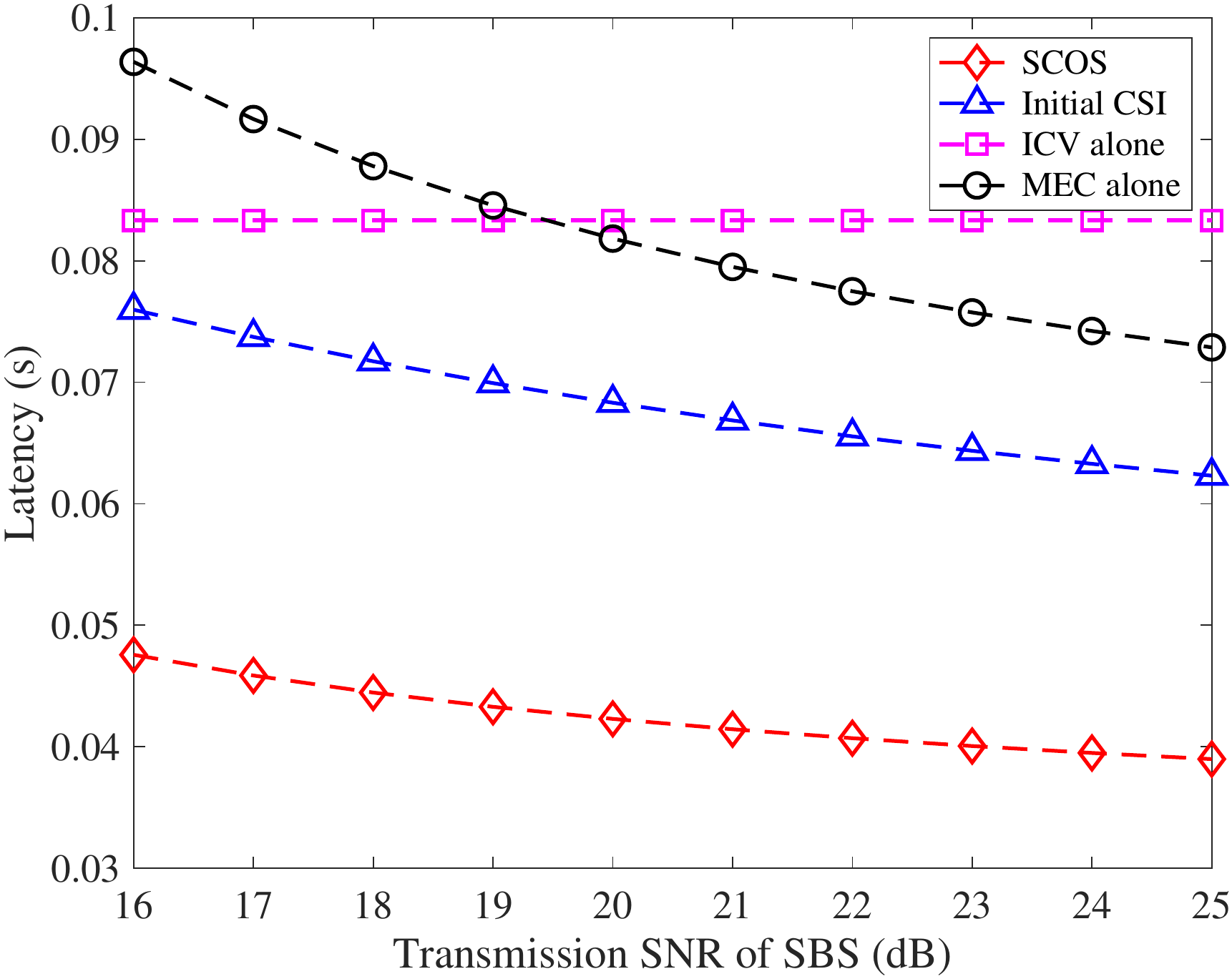}
\caption{Latency in terms of transmission SNR of SBS, where we set $a_m=12\times 10^7$ bits/s, $a_v=6\times 10^7$ bit/s, $N_m=10$, $N_e=4$, and transmission SNR of ICV is 10 dB.}\label{sim3}
\end{minipage}
\hspace{0.2cm}
\begin{minipage}[t]{0.47\linewidth}
\centering
\includegraphics[width=1\linewidth]{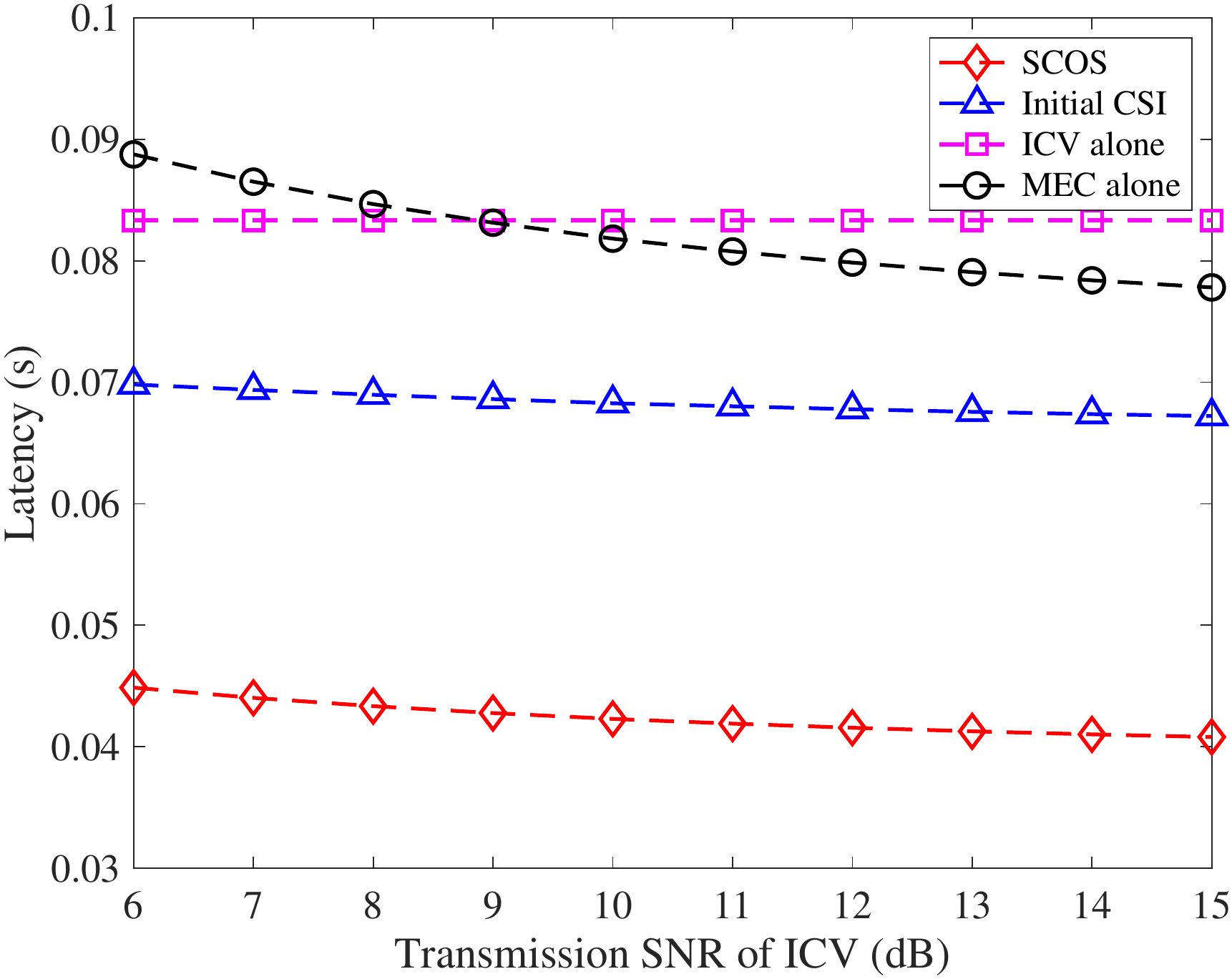}
\caption{Latency in terms of transmission SNR of ICV, where we set $a_m=12\times 10^7$ bits/s, $a_v=6\times 10^7$ bit/s, $N_m=10$, $N_e=4$, and transmission SNR of MEC server is 20 dB.}\label{sim4}
\end{minipage}
\end{figure*}

\begin{figure*}[h]
\begin{minipage}[t]{0.47\linewidth}
\centering
\includegraphics[width=1\linewidth]{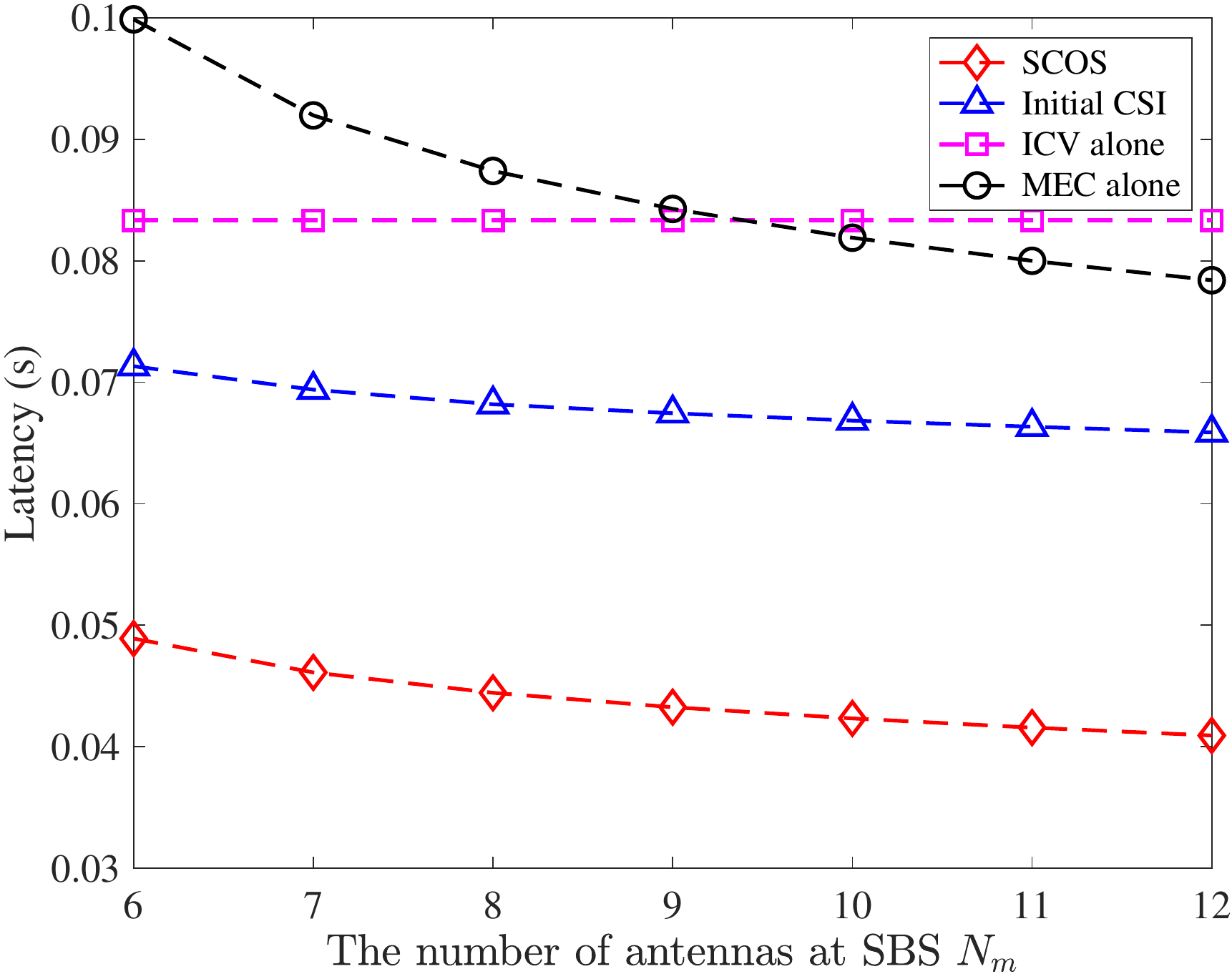}
\caption{Latency in terms of the number of antennas at SBS, i.e., $N_m$, where we set $a_m=12\times 10^7$ bit/s, $a_v=6\times 10^7$ bit/s, $N_e=4$, and transmission SNRs of SBS and ICV are 20 dB and 10 dB, respectively.}\label{sim5}
\end{minipage}
\hspace{0.2cm}
\begin{minipage}[t]{0.47\linewidth}
\centering
\includegraphics[width=1\linewidth]{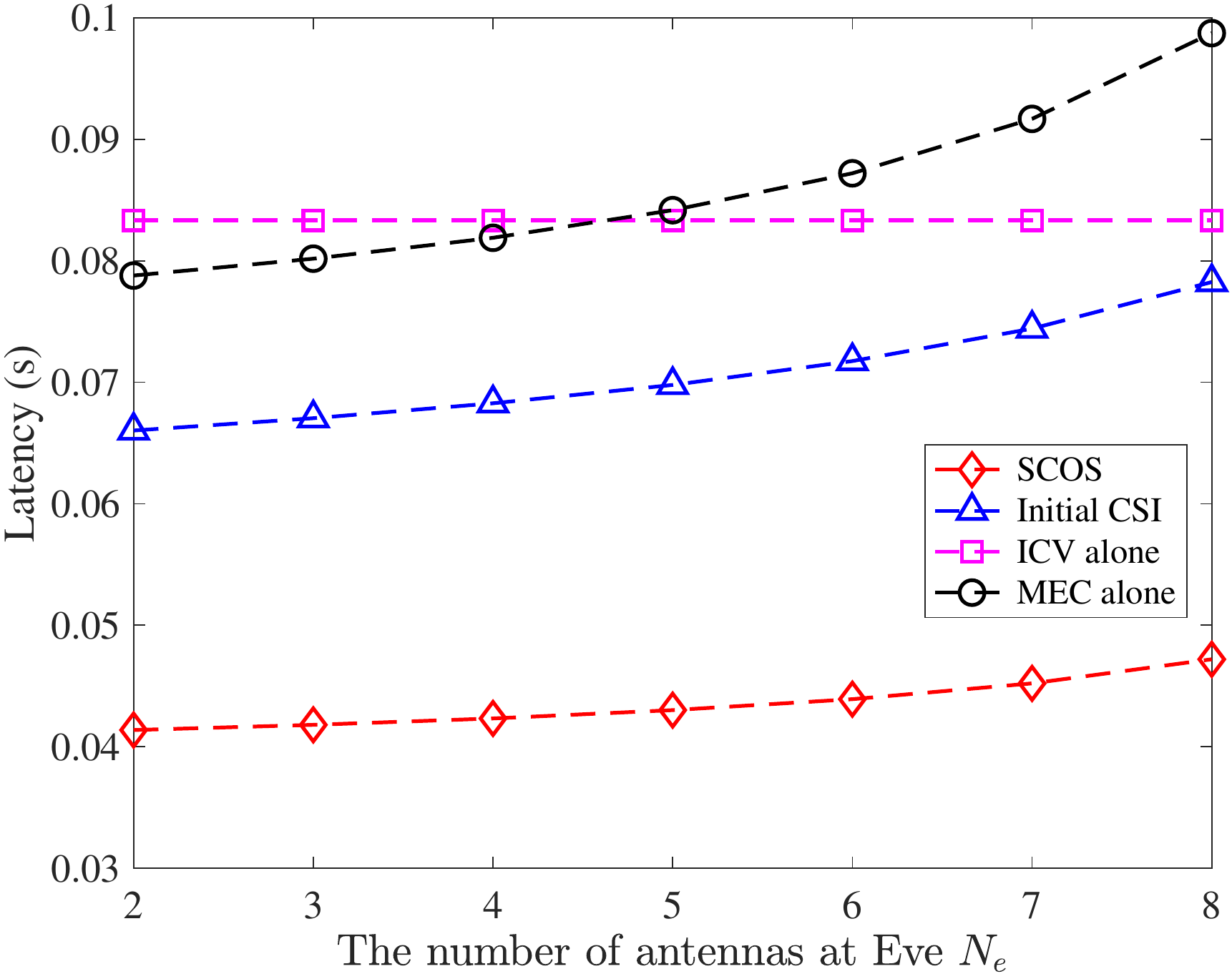}
\caption{Latency in terms of the number of antennas at Eve, i.e., $N_e$, where we set $a_m=12\times 10^7$ bit/s, $a_v=6\times 10^7$ bit/s, $N_m=10$, and transmission SNRs of SBS and ICV are 20 dB and 10 dB, respectively.}\label{sim6}
\end{minipage}
\end{figure*}

Theoretical results of Propositions 4 and 5 are verified in Figs. \ref{efsru} and \ref{efsrd}, respectively, where Monte Carlo simulations were done based on Eqns. (\ref{effsr}) and (\ref{outaged}), respectively. Here, we assume that the main capacities of uplink and downlink are 6 and 5 bit/s/Hz, respectively. The main capacities of uplink are higher than that of downlink channels, which is reasonable because the uplink transmission phases have the access to additional power offered by ICV. As shown in Fig. \ref{efsru}, we find that effective secrecy rate $\hat{R}_u$ increases with an increasing wiretap code rate $R_u$ at the beginning, since instantaneous secrecy rate $R_1$ is larger than $R_u$ with a large probability. Then, $\hat{R}_u$ will decrease with $R_u$ because a large $R_u$ will cause a large secrecy outage probability. The downlink phase in Fig. \ref{efsrd} shows a similar phenomenon in the uplink phase. Moreover, we find only one peak in each curve in Figs. \ref{efsru} and \ref{efsrd}, which is consistent with the results shown in \cite{Zhou2011RethinkingSecrecyOutage}, meaning that we can use unimodal function-aimed search algorithms, such as golden-section search.

The simulation results are provided to investigate joint impacts of computing capacities, transmission SNR, and the number of antennas on system latency. We assume that an autonomous control task has 610 KB images, in which these images are processed with full granular data-partition \cite{Wang2016MobileEdgeComputing, Chen2016EfficientMultiUser, Mao2017SurveyMobileEdge}, and part of the data ($\eta^*\times 610$ KB) will be compressed and uploaded. We use 20 MHz bandwidth\footnote{3GPP Release 16 supports 10, 20, 30, and 40 MHz bandwidth for New Radio (NR)-V2X \cite{3gpp3}, and will introduce new channel bandwidth for NR-V2X licensed bands for future applications.} for uplink and downlink as defined in 3GPP LTE-V2X \cite{3gpp3} \cite{Molina-Masegosa2017LTEVSidelink}. The compressed ratio of a file $\beta$ is set to 0.4 \cite{Taubman2000Highperformancescalable}. The output to input data ratio in the MEC server $\alpha$ is set to 0.4. In addition, four different schemes, i.e., ICV alone, MEC server alone, SCOS, and the scheme that tasks are partitioned with CSIs obtained at the beginning of transmissions, are compared, which are described as follows.
\begin{enumerate} 
\item ICV alone: The whole task is executed in an ICV.
\item MEC alone: The whole task is uploaded and executed in an MEC server. PHY-layer security schemes with effective uplink and downlink secrecy rate maximization are used.
\item SCOS: Simulations use ergodic secrecy rates for task partition, and use PHY-layer security schemes with effective secrecy rate maximization in both uplink and downlink.
\item Task partition with initial CSIs: Simulations use CSIs obtained at the beginning of uplink transmissions for task partition and also use PHY-layer security methods. That is, estimate $\mathbf{h}_{vm}$ before sending data to an SBS, with an assumption of $\mathbf{h}_{mv}=\mathbf{h}_{vm}$, which are constant during the entire process. We also assume that Eve's CSIs can be obtained, and then we calculate the corresponding secrecy rates of uplink and downlink phases for task partition. The assumption of constant initial CSIs was also used in the literature, such as \cite{HuEarlyAccess2019, Cheng2019, Wang2016MobileEdgeComputing, Chen2016EfficientMultiUser}.
\end{enumerate}

In the simulations, we ignored the overhead of data compression/decompression and task partition because these customized functions can be decoupled from the data forwarding via software-defined network (SDN) technologies\cite{Yan2016, Shu2016}, in which the overhead is much smaller than that of vehicular computation. Each simulation runs 10000 times.

Next, we want to show the impact of MEC server computing capacity on system latency in Fig. \ref{sim1}. Several observations can be made as follows. First, SCOS has a better performance than others, and the scheme of MEC server alone outperforms the scheme with initial CSIs when MEC server computing capacity is large enough. Second, latency decreases with an increasing MEC server computing capacity, but its gain is not large enough because the smallest one of $a_m$, $R_u$, and $R_d$ decides the capability of edge computing as discussed in Remark 4. Unilateral increasing of $a_m$ can not provide a large gain if $R_u$ and $R_d$ are limited. Also, if we use local computing resources only, latency keeps constant. Note that latency of simulations is approximately 50 ms with the proposed scheme. Nevertheless, the round-trip time of LTE BS caused by access and control scheduling should be added in total latency in real-world systems. The round-trip time is approximately 20 ms in 4G and will be reduced to less than 10 ms in 5G NR \cite{Parvez2018}.

The effect of ICV computing capacity is shown in Fig. \ref{sim2}. Similar to the MEC server computing capacity, an increasing ICV computing capacity also reduces latency, and yet provides a higher gain than the MEC server computing capacity as the lines drop quickly. It shows that a better way to improve vehicle performance is to increase ICV computing capacity, rather than the MEC server, because MEC-assisted vehicular computation follows the cask principle.

The effects of transmission SNRs of SBS and ICV are examined in Figs. \ref{sim3} and \ref{sim4}, respectively. These figures show similar trends that latency reduces with an increasing transmission SNR in SCOS, MEC server alone, and the scheme with initial CSIs, because by increasing transmission SNR of SBS or ICV, secrecy rate between SBS and ICV increases, and latency of MEC-assisted schemes reduces. As shown in Fig. \ref{sim3}, we can observe that the gap between SCOS and the scheme with initial CSIs is small at the beginning, and then the gap will be enlarged with an increasing SNR of SBS. It means that the scheme with initial CSIs is very sensitive to the transmission power of SBS as SBS undertakes the tasks of secure transmissions. 

Finally, Figs. \ref{sim5} and \ref{sim6} illustrate the impacts of the number of antennas on system latency. From Fig. \ref{sim5}, we can observe that an increasing number of antennas at Eve will increase latency because secrecy rates will be reduced with an increasing number of antennas at Eve. An opposite trend can be seen in Fig. \ref{sim6}, where secrecy rates increase with an increasing number of antennas at SBS, such that latency will be decreased. Also, the gain of $N_m$ is small, because an increasing $N_m$ provides a small gain on $R_u$ that is limited by the upper bound $C_{v,u}$. Nevertheless, an increasing number of antennas at Eve will cause a large latency in the system.

\section{CONCLUSIONS AND FUTURE WORKS}\label{conclusion}
In this paper, we proposed SCOS, in which an ICV can offload part of computation tasks to an MEC server to minimize computation delay for latency-critical vehicular communications. Specifically, we designed a computation task partition scheme, as well as a way to secure uplink and downlink transmissions between ICV and SBS. We adopted ergodic uplink and downlink secrecy rates for task partition and adaptive wiretap coding to avoid a large secrecy outage probability incurred by high mobility. Simulation results have shown that SCOS can reduce system latency significantly by almost 40\% in comparison with state-of-the-art schemes. According to the computation complexity analysis, the proposed scheme can be applied to computation offloading in 5G for autonomous driving applications because it does not cause a huge burden to 5G wireless communication systems. As one of our future works, we will integrate multi-antenna technologies with ICV, in which multiple independent secrecy information streams will be transmitted between ICV and SBS, so that feedback latency of computation tasks can be reduced further. Also, we should address the problems caused by the correlation between wiretap and legitimate channels if Eve has mobility like ICV.

\vspace{0.1in}
\section*{APPENDIX}

\subsection{Proof of Proposition 1}\label{propro1}
Recall Eqn. (\ref{luesc}) as 
\begin{flalign}\label{rcluesc}
\bar{R}_{u}= \text{E}_{\mathbf{h}_{vm}}[C_{v,u}]- \text{E}_{\mathbf{h}_{vm},\mathbf{h}_{ve},\mathbf{H}_{me}}[C_{e,u}].
\end{flalign}
Based on \cite[Eq. (21)]{Shin2003}, we know $\text{E}_{\mathbf{h}_{vm}}[C_{v,u}]$ has a closed-form expression as
\begin{flalign}\label{mainescup}
\text{E}_{\mathbf{h}_{vm}}[C_{v,d}]=\frac{B_1}{\ln(2)}\exp(\rho_1)\sum_{k=0}^{N_m-1}E_{k+1}(\rho_1),
\end{flalign}
where $\rho_1 =\frac{\sigma_{vm}^2}{P_v}$, and $E_{\tau(z)}$ is an exponential integral of order $\tau$ as defined in Eqn. (\ref{EN}).

Then, we can simplify $C_{e,u}$ as
\begin{flalign}
C_{e,u}&=B_1\log_2\det\bigg(\mathbf{I}_{N_e}+\frac{P_v\mathbf{h}_{ve}\mathbf{h}_{ve}^{\dagger}}{\sigma_{ve}^2\mathbf{I}_{N_e}+\frac{P_m}{N_m-1}\mathbf{H}_{1}\mathbf{H}_{1}^{\dagger}}\bigg) \\ \notag
&=B_1\log_2\det\bigg(\frac{\mathbf{I}_{N_e}+\rho_2\mathbf{H}_{1}\mathbf{H}_{1}^{\dagger}+P_v/\sigma_{ve}^2\mathbf{h}_{ve}\mathbf{h}_{ve}^{\dagger}}{\mathbf{I}_{N_e}+\rho_2\mathbf{H}_{1}\mathbf{H}_{1}^{\dagger}}\bigg) \\ \notag
&=B_1\log_2\det\bigg(\frac{\mathbf{I}_{N_e}+\mathbf{H}_{2}\mathbf{Q}\mathbf{H}_{2}^{\dagger}}{\mathbf{I}_{N_e}+\rho_2\mathbf{H}_{1}\mathbf{H}_{1}^{\dagger}}\bigg),
\end{flalign}
where $\mathbf{Q}=\text{diag}(P_v/\sigma_{ve}^2,\rho_2,...,\rho_2)$, $\mathbf{H}_1=\mathbf{H}_{me}\mathbf{G}_u\in\mathbb{C}^{N_e\times (N_m-1)}$, and $\mathbf{H}_2=[\mathbf{h}_{ve}, \mathbf{H}_{1}]\in\mathbb{C}^{N_e\times N_m}$. Hence, the second part of Eqn. (\ref{rcluesc}), i.e., $\text{E}_{\mathbf{h}_{vm},\mathbf{h}_{ve},\mathbf{H}_{me}}[C_{e,u}]$, can be expressed as
\begin{flalign}
&\text{E}_{\mathbf{h}_{vm},\mathbf{h}_{ve},\mathbf{H}_{me}}[C_{e,u}]\notag \\
&=\text{E}\bigg[B_1\log_2\det\bigg(\frac{\mathbf{I}_{N_e}+\mathbf{H}_{2}\mathbf{Q}\mathbf{H}_{2}^{\dagger}}{\mathbf{I}_{N_e}+\rho_2\mathbf{H}_{1}\mathbf{H}_{1}^{\dagger}}\bigg)\bigg]\notag \\
&=B_1\{\Psi(\mathbf{H}_2,\rho_2,P_v)-C(\mathbf{H}_1,\rho_2)\},
\end{flalign}
where
\begin{subequations}
\begin{flalign}
&\Psi(\mathbf{H}_2,\rho_2,P_v)=\text{E}_{\mathbf{H}_2}[\log_2\det(\mathbf{I}_{N_e}+\mathbf{H}_{2}\mathbf{Q}\mathbf{H}_{2}^{\dagger})], \\
&C(\mathbf{H}_1, \rho_2)=\text{E}_{\mathbf{H}_1}[\log_2\det(\mathbf{I}_{N_e}+\rho_2\mathbf{H}_{1}\mathbf{H}_{1}^{\dagger})],
\end{flalign}
\end{subequations}
because $\mathbf{H}_1$ and $\mathbf{H}_2$ are mutually independent complex Gaussian random matrices \cite{Liu2017SecrecyCapacityAnalysis}. According to \cite[Eq. (18)]{Shin2003}, we have a closed-form expression of $C(\mathbf{H}_1, \rho_2)$ as in Eqn. (\ref{erCm1}). $\Psi(\mathbf{H}_2,\rho_2,P_v)$ can be deduced by ergodic mutual information in an MIMO Rayleigh fading channel with an input covariance matrix $\mathbf{Q}$, which is written as
\begin{flalign}\label{psiemi}
&\Psi(\mathbf{H}_2,\rho_2,P_v)=C_{\text{SU}}(N_m,P_v,\mathbf{Q}),
\end{flalign}
where the ergodic mutual information $C_{\text{SU}}(N_m,P_v,\mathbf{Q})$ is formulated in \cite[Eq. (30)] {Chiani2010}.

Substituting $C(\mathbf{H}_1, \rho_2)$, (\ref{psiemi}), and (\ref{mainescup}) to Eqn. (\ref{rcluesc}), we get Eqn. (\ref{ESRu}) in Proposition 1. The proof is completed. \hfill $\IEEEQEDclosed$

\subsection{Proof of Corollary 1}\label{procor1}
Based on \cite[Eq. (80)]{McKay2005} and the condition $N_e\leq N_m-1$, we have the lower bound of $C(\mathbf{H}_1, \rho_2)$ as 
\begin{flalign}\label{lbc1}
&C(\mathbf{H}_1, \rho_2)\geq N_e\log_2(\rho_2)+\sum_{i=0}^{N_e-1}\psi(N_m-1-i),
\end{flalign}
where $\psi(x)$ is defined in Eqn. (\ref{psi}).

Then, we can rewrite $\Psi(\mathbf{H}_2,\rho_2,P_v)$ as
\begin{flalign}
\Psi(\mathbf{H}_2,\rho_2,P_v)&=\text{E}_{\mathbf{H}_2}[\log_2\det(\mathbf{I}_{N_e}+\mathbf{H}_{2}\mathbf{Q}\mathbf{H}_{2}^{\dagger})] \notag \\
&=\text{E}_{\mathbf{H}_2}[\log_2\det(\mathbf{I}_{N_e}+\rho_2\mathbf{H}_{2}\mathbf{Q}'\mathbf{H}_{2}^{\dagger})],
\end{flalign}
where 
\begin{flalign}
\mathbf{Q}'=\text{diag}\Big(\frac{P_v}{\sigma_{ve}^2\rho_2},1,...,1\Big).
\end{flalign}
We use Jensen's inequality to get
\begin{flalign}
&\text{E}_{\mathbf{H}_2}[\log_2\det(\mathbf{I}_{N_e}+\rho_2\mathbf{H}_{2}\mathbf{Q}'\mathbf{H}_{2}^{\dagger})]  \notag \\
& \leq \log_2\{\det(\mathbf{I}_{N_e}+\rho_2\text{E}_{\mathbf{H}_2}[\mathbf{H}_{2}\mathbf{Q}'\mathbf{H}_{2}^{\dagger}])\}.
\end{flalign}

Based on \cite[Eqs. (5) and (7)]{Salo2006}, in a high SNR region, we have the upper bound of $\Psi(\mathbf{H}_2,\rho_2,P_v)$ as 
\begin{flalign}\label{uppsi}
& \Psi(\mathbf{H}_2,\rho_2,P_v) \notag \\
& \leq \log_2\Big\{\rho_2^{N_e}\sum_{i=0}^{N_e}[N_m-i+1]_i\text{Tr}_{i}(\mathbf{Q}') \Big\} \notag \\ 
&=N_e\log_2(\rho_2)+\log_2\Big\{\sum_{i=0}^{N_e}[N_m-i+1]_i\text{Tr}_{i}(\mathbf{Q}') \Big\},
\end{flalign}
where the function $\text{Tr}_{i}(\mathbf{W}), i=0,...,K$ is the $i$-th elementary symmetric function of $(K \times K)$ matrix $\mathbf{W}$. $\text{Tr}_{i}(\mathbf{W})$ depends only on the eigenvalues of $\mathbf{W}$, which are denoted by $\lambda_1,...,\lambda_K$. For instance, $\text{Tr}_{0}(\mathbf{W})=1$, $\text{Tr}_{1}(\mathbf{W})=\text{Tr}(\mathbf{W})$, and $\text{Tr}_{K}(\mathbf{W})=\det(\mathbf{W})$. In general, $\text{Tr}_{i}(\mathbf{W})=\sum_{i=1}^K \lambda_{j_1}\times\lambda_{j_2}...\times\lambda_{j_i}$, where the sum is calculated over all $\binom{K}{i}$ combinations of $i$ indices with $j_1<...<j_i$. Based on the characteristics of $\mathbf{Q}'$, we have $\text{Tr}_{0}(\mathbf{Q}')=1$, $\text{Tr}_{1}(\mathbf{Q}')=P_v/(\sigma_{ve}^2\rho_2)+N_m-1$, $\text{Tr}_{N_m}(\mathbf{Q}')=P_v/(\sigma_{ve}^2\rho_2)$, and 
\begin{flalign}
&\sum_{i=0}^{N_e}[N_m-i+1]_i\text{Tr}_{i}(\mathbf{Q}') \notag \\
&=1+\sum_{i=1}^{N_e}[N_m-i+1]_i\binom{N_m-1}{i-1}\frac{P_v}{\sigma_{ve}^2\rho_2} \notag \\  
&+\sum_{i=1}^{N_e}\sum_{j=2}^{N_m}[N_m-i+1]_i\binom{N_m-j}{i-1},
\end{flalign}
which reduces the complexity of the trace operations. Substituting (\ref{lbc1}), (\ref{uppsi}), and (\ref{mainescup}) to Eqn. (\ref{rcluesc}), we obtain the lower bound of Eqn. (\ref{rcluesc}) as shown in Eqn. (\ref{lbescu}) of Corollary 1. The proof is completed. \hfill $\IEEEQEDclosed$

\subsection{Proof of Corollary 2}\label{procor2}
Recall the ergodic secrecy rate in downlink phases $\bar{R}_d$ as 
\begin{flalign}\label{reESR1}
\bar{R}_d  = & B_2 \{\Phi(\rho_3)+ C(\mathbf{H}_3,\rho_4) -  C(\mathbf{H}_4,\rho_4)\}.
\end{flalign}
Similar to Corollary 1, the lower bound of $C(\mathbf{H}_3, \rho_4)$ in a high SNR region can be expressed as  
\begin{flalign}\label{lbc2}
&C(\mathbf{H}_3, \rho_4)\geq N_e\log_2(\rho_4)+\sum_{i=0}^{N_e-1}\psi(N_m-1-i),
\end{flalign}
where $\psi(x)$ is defined in Eqn. (\ref{psi}). Based on \cite[Eq. (81)]{McKay2005}, we have the upper bound of $C(\mathbf{H}_4, \rho_4)$ in a high SNR region as 
\begin{flalign}\label{lbc3}
&C(\mathbf{H}_4, \rho_4)\leq N_e\log_2(\rho_4)+\log_2\bigg(\frac{N_m!}{(N_m-N_e)!}\bigg).
\end{flalign}

Substituting Eqns. (\ref{lbc2}), (\ref{lbc3}), and (\ref{mainescup}) to Eqn. (\ref{reESR1}), we have Eqn. (\ref{lbdec}) in Corollary 2. The proof is completed. \hfill $\IEEEQEDclosed$

\subsection{Proof of Proposition 3}\label{propro3}
We can adopt a reverse-proof method to show Proposition 3. Differentiating $T_{v}(\eta)$ and $T_{\text{MEC}}(\eta)$ with respect to $\eta$, we get
\begin{subequations}
\begin{flalign}
&\frac{dT_{v}(\eta)}{d\eta}=\frac{M}{a_v},\notag \\ 
&\frac{dT_{\text{MEC}}(\eta)}{d\eta}=-\big(\frac{M}{a_m}+\frac{\beta M}{\bar{R}_u}+\frac{\alpha M}{\bar{R}_d}\big),
\end{flalign}
\end{subequations}
which means that $T_{v}(\eta)$ is a monotonically increasing function of $\eta$, and $T_{\text{MEC}}(\eta)$ is a monotonically decreasing function of $\eta$. 

Find $\eta_0$ as 
\begin{flalign}
f_0=T_{v}(\eta_0)=T_{\text{MEC}}(\eta_0),
\end{flalign}
and assume that $\eta_1\neq \eta_0$ minimizes $\max\{T_{v}(\eta),T_{\text{MEC}}(\eta)\}$ as
\begin{flalign}
f_1=\max\big(T_{v}(\eta_1),T_{\text{MEC}}(\eta_1)\big),
\end{flalign}
such that $f_1< f_0$. First, considering the case $T_{v}(\eta_1)>T_{\text{MEC}}(\eta_1)$, we have
\begin{flalign}\label{p32}
T_{\text{MEC}}(\eta_0)=T_{v}(\eta_0)> T_{v}(\eta_1)>T_{\text{MEC}}(\eta_1).
\end{flalign}
Since $T_{v}(\eta_0)> T_{v}(\eta_1)$, we get $\eta_0> \eta_1$. In this case, $T_{\text{MEC}}(\eta_0)< T_{\text{MEC}}(\eta_1)$ because $T_{\text{MEC}}(\eta)$ is a monotonically decreasing function, which is contradictory to Eqn. (\ref{p32}). For $T_{v}(\eta_1)\leq T_{\text{MEC}}(\eta_1)$, the proof is similar to the case $T_{v}(\eta_1)>T_{\text{MEC}}(\eta_1)$. Thus, $\eta_1\neq \eta_0$ can not minimize $\max\{T_{v}(\eta),T_{\text{MEC}}(\eta)\}$, and only $\eta_0$ can minimize it. 

Then, we solve the equation
\begin{flalign}\label{p31}
T_{\text{MEC}}(\eta)=T_{v}(\eta),
\end{flalign}
and get the solution as in Eqn. (\ref{oall}). The proof is completed. \hfill $\IEEEQEDclosed$

\subsection{Proof of Proposition 4}\label{propro4}
Here, Lemma 1 is used to prove Proposition 4.
\begin{lemma}
[Proved in \cite{Gao1998TheoreticalreliabilityMMSE}] For $a\times 1$ vector $\mathbf{h}$ and $a\times(b-1)$ matrix $\mathbf{H}$ that consist of i.i.d. complex Gaussian entries obeying $\mathcal{CN}(0,1)$, the complementary cumulative distribution function (CCDF) of $Z=\mathbf{h}^{\dagger}(r\mathbf{I}_a+\mathbf{H}\mathbf{H}^{\dagger})^{-1}\mathbf{h}$ is given by 
\begin{flalign}
&F_Z(z)=\exp(-zr)\sum_{k=0}^{a-1}\frac{A_k(z)}{k!}(zr)^k, \notag \\
& A_k(z)=\frac{\sum_{n=0}^{a-k-1}\binom{b-1}{n}z^n}{(1+z)^{b-1}},
\end{flalign}
where $r$ is a non-negative real number.
\end{lemma}

As channel capacity $C_{v,u}$ can be calculated by $\mathbf{h}_{vm}$ and Eqn. (6a), and $\mathbf{H}_1=\mathbf{H}_{me}\mathbf{G}_u$ is a cyclic symmetry complex Gaussian matrix \cite{Liu2017SecrecyCapacityAnalysis}, we can transform $P_{\text{out}}(R_u)$ to
\begin{flalign}\label{probability}
P_{\text{out}}(R_u)&= P(C_{e,u}> C_{v,u}-R_u) \notag \\
&=P(\mathbf{h}_{ve}^{\dagger}(a_1\mathbf{I}_{N_e}+\mathbf{H}_{1}\mathbf{H}_{1}^{\dagger})^{-1}\mathbf{h}_{ve}\geq \phi_1) \notag \\
&=F_{Z_1}(\phi_1),
\end{flalign}
where $Z_1=\mathbf{h}_{ve}^{\dagger}(a_1\mathbf{I}_{N_e}+\mathbf{H}_{1}\mathbf{H}_{1}^{\dagger})^{-1}\mathbf{h}_{ve}$, $\phi_1=\frac{P_m}{P_v(N_m-1)} (2^{C_{v,u}-R_u}-1)$, and $a_1=\frac{\sigma_{ve}^2(N_m-1)}{P_m}$. Substituting Eqn. (\ref{probability}) into Eqn. (\ref{effsr}), we obtain the expression of effective secrecy rates as in Eqn. (\ref{outage}). The proof is completed. \hfill $\IEEEQEDclosed$

\vspace{0.15in}
\bibliographystyle{IEEEtran}
\bibliography{references}

\begin{thebibliography}{10}
\providecommand{\url}[1]{#1}
\csname url@samestyle\endcsname
\providecommand{\newblock}{\relax}
\providecommand{\bibinfo}[2]{#2}
\providecommand{\BIBentrySTDinterwordspacing}{\spaceskip=0pt\relax}
\providecommand{\BIBentryALTinterwordstretchfactor}{4}
\providecommand{\BIBentryALTinterwordspacing}{\spaceskip=\fontdimen2\font plus
\BIBentryALTinterwordstretchfactor\fontdimen3\font minus
  \fontdimen4\font\relax}
\providecommand{\BIBforeignlanguage}[2]{{%
\expandafter\ifx\csname l@#1\endcsname\relax
\typeout{** WARNING: IEEEtran.bst: No hyphenation pattern has been}%
\typeout{** loaded for the language `#1'. Using the pattern for}%
\typeout{** the default language instead.}%
\else
\language=\csname l@#1\endcsname
\fi
#2}}
\providecommand{\BIBdecl}{\relax}
\BIBdecl

\bibitem{Zhang2018}
S.~{Zhang}, J.~{Chen}, F.~{Lyu}, N.~{Cheng}, W.~{Shi}, and X.~{Shen},
  ``Vehicular communication networks in the automated driving era,'' \emph{IEEE
  Commun. Mag.}, vol.~56, no.~9, pp. 26--32, Sep. 2018.

\bibitem{reliable-V2V}
F.~Lyu, H.~Zhu, N.~Cheng, H.~Zhou, W.~Xu, M.~Li, and X.~Shen, ``{Characterizing
  Urban Vehicle-to-Vehicle Communications for Reliable Safety Applications},''
  \emph{IEEE Trans. Intell. Transp. Syst.}, vol.~21, no.~6, pp. 2586--2602,
  Jun. 2019.

\bibitem{Zhou2017}
J.~{Zhou}, Z.~{Cao}, X.~{Dong}, and A.~V. {Vasilakos}, ``Security and privacy
  for cloud-based {IoT}: Challenges,'' \emph{IEEE Commun. Mag.}, vol.~55,
  no.~1, pp. 26--33, Jan. 2017.

\bibitem{Yang2018Intelligentconnectedvehicles}
D.~Yang, K.~Jiang, D.~Zhao, C.~Yu, Z.~Cao, S.~Xie, Z.~Xiao, X.~Jiao, S.~Wang,
  and K.~Zhang, ``Intelligent and connected vehicles: {C}urrent status and
  future perspectives,'' \emph{Science China Technological Sciences}, vol.~61,
  no.~10, pp. 1446--1471, Oct. 2018.

\bibitem{Mei2018}
J.~{Mei}, K.~{Zheng}, L.~{Zhao}, Y.~{Teng}, and X.~{Wang}, ``A latency and
  reliability guaranteed resource allocation scheme for {LTE} {V2V}
  communication systems,'' \emph{IEEE Trans. Wireless Commun.}, vol.~17, no.~6,
  pp. 3850--3860, Jun. 2018.

\bibitem{Tesla}
\BIBentryALTinterwordspacing
Tesla online store. [Online]. Available:
  \url{https://www.tesla.com/models/design\#autopilot}
\BIBentrySTDinterwordspacing

\bibitem{Mao2017SurveyMobileEdge}
Y.~{Mao}, C.~{You}, J.~{Zhang}, K.~{Huang}, and K.~B. {Letaief}, ``A survey on
  mobile edge computing: {The} communication perspective,'' \emph{IEEE Commun.
  Surveys Tuts.}, vol.~19, no.~4, pp. 2322--2358, Fourthquarter 2017.

\bibitem{HuEarlyAccess2019}
X.~{Hu}, K.~{Wong}, K.~{Yang}, and Z.~{Zheng}, ``{UAV}-assisted relaying and
  edge computing: {S}cheduling and trajectory optimization,'' \emph{IEEE Trans.
  Wireless Commun.}, vol.~18, no.~10, pp. 4738--4752, Oct. 2019.

\bibitem{Cheng2019}
N.~{Cheng}, F.~{Lyu}, W.~{Quan}, C.~{Zhou}, H.~{He}, W.~{Shi}, and X.~{Shen},
  ``Space/aerial-assisted computing offloading for {IoT} applications: A
  learning-based approach,'' \emph{IEEE J. Sel. Areas Commun.}, vol.~37, no.~5,
  pp. 1117--1129, May 2019.

\bibitem{Wang2016MobileEdgeComputing}
Y.~{Wang}, M.~{Sheng}, X.~{Wang}, L.~{Wang}, and J.~{Li}, ``Mobile-edge
  computing: Partial computation offloading using dynamic voltage scaling,''
  \emph{IEEE Trans. Commun.}, vol.~64, no.~10, pp. 4268--4282, Oct. 2016.

\bibitem{Chen2016EfficientMultiUser}
X.~{Chen}, L.~{Jiao}, W.~{Li}, and X.~{Fu}, ``Efficient multi-user computation
  offloading for mobile-edge cloud computing,'' \emph{IEEE/ACM Trans. Netw.},
  vol.~24, no.~5, pp. 2795--2808, Oct. 2016.

\bibitem{Yang2018SmallCellAssisted}
X.~{Yang}, X.~{Wang}, Y.~{Wu}, L.~P. {Qian}, W.~{Lu}, and H.~{Zhou},
  ``Small-cell assisted secure traffic offloading for narrowband internet of
  thing ({NB-IoT}) systems,'' \emph{IEEE Internet Things J.}, vol.~5, no.~3,
  pp. 1516--1526, Jun. 2018.

\bibitem{Xu2019a}
J.~{Xu} and J.~{Yao}, ``Exploiting physical-layer security for multiuser
  multicarrier computation offloading,'' \emph{IEEE Wireless Commun. Lett.},
  vol.~8, no.~1, pp. 9--12, Feb. 2019.

\bibitem{Bai2019}
T.~{Bai}, J.~{Wang}, Y.~{Ren}, and L.~{Hanzo}, ``Energy-efficient computation
  offloading for secure {UAV}-edge-computing systems,'' \emph{IEEE Trans. Veh.
  Technol.}, vol.~68, no.~6, pp. 6074--6087, Jun. 2019.

\bibitem{Yan2016}
Z.~Yan, P.~Zhang, and A.~V. Vasilakos, ``A security and trust framework for
  virtualized networks and software-defined networking,'' \emph{Security and
  Communication Networks}, vol.~9, no.~16, pp. 3059--3069, Nov. 2016.

\bibitem{Brecht2018SecurityCredentialManagement}
B.~{Brecht}, D.~{Therriault}, A.~{Weimerskirch}, W.~{Whyte}, V.~{Kumar},
  T.~{Hehn}, and R.~{Goudy}, ``A security credential management system for
  {V2X} communications,'' \emph{IEEE Trans. Intell. Transp. Syst.}, vol.~19,
  no.~12, pp. 3850--3871, Dec. 2018.

\bibitem{Zhou2015SecurePrivacyPreserving}
J.~{Zhou}, X.~{Dong}, Z.~{Cao}, and A.~V. {Vasilakos}, ``Secure and privacy
  preserving protocol for cloud-based vehicular {DTNs},'' \emph{IEEE Trans.
  Inf. Forensics Security}, vol.~10, no.~6, pp. 1299--1314, Jun. 2015.

\bibitem{Zhou2013}
J.~{Zhou}, Z.~{Cao}, X.~{Dong}, X.~{Lin}, and A.~V. {Vasilakos}, ``Securing
  m-healthcare social networks: {C}hallenges, countermeasures and future
  directions,'' \emph{IEEE Wireless Commun.}, vol.~20, no.~4, pp. 12--21, Aug.
  2013.

\bibitem{Zhang2018HealthDepEfficientSecure}
Y.~{Zhang}, C.~{Xu}, H.~{Li}, K.~{Yang}, J.~{Zhou}, and X.~{Lin},
  ``Health{D}ep: An efficient and secure deduplication scheme for
  cloud-assisted {eHealth} systems,'' \emph{IEEE Trans. Ind. Informat.},
  vol.~14, no.~9, pp. 4101--4112, Sep. 2018.

\bibitem{Wazid2020}
M.~Wazid, A.~K. Das, V.~{Bhat K}, and A.~V. Vasilakos, ``{LAM-CIoT}:
  Lightweight authentication mechanism in cloud-based iot environment,''
  \emph{Journal of Network and Computer Applications}, vol. 150, pp. 1--16,
  Jan. 2020.

\bibitem{Wang2018PhysicalLayerSecurity}
W.~{Wang}, K.~C. {Teh}, S.~{Luo}, and K.~H. {Li}, ``Physical layer security in
  heterogeneous networks with pilot attack: {A} stochastic geometry approach,''
  \emph{IEEE Trans. Commun.}, vol.~66, no.~12, pp. 6437--6449, Dec. 2018.

\bibitem{Zhou2011RethinkingSecrecyOutage}
X.~{Zhou}, M.~R. {McKay}, B.~{Maham}, and A.~{Hjorungnes}, ``Rethinking the
  secrecy outage formulation: {A} secure transmission design perspective,''
  \emph{IEEE Commun. Lett.}, vol.~15, no.~3, pp. 302--304, Mar. 2011.

\bibitem{Harrison2019}
W.~K. {Harrison}, T.~{Fernandes}, M.~A.~C. {Gomes}, and J.~P. {Vilela},
  ``Generating a binary symmetric channel for wiretap codes,'' \emph{IEEE
  Trans. Inf. Forensics Security}, vol.~14, no.~8, pp. 2128--2138, Aug. 2019.

\bibitem{Oggier2016}
F.~{Oggier}, P.~{Sole}, and J.~{Belfiore}, ``Lattice codes for the wiretap
  gaussian channel: Construction and analysis,'' \emph{IEEE Trans. Inf.
  Theory}, vol.~62, no.~10, pp. 5690--5708, Oct. 2016.

\bibitem{Mahdavifar2011}
H.~{Mahdavifar} and A.~{Vardy}, ``Achieving the secrecy capacity of wiretap
  channels using polar codes,'' \emph{IEEE Trans. Inf. Theory}, vol.~57,
  no.~10, pp. 6428--6443, Oct. 2011.

\bibitem{Sardellitti2015JointOptimizationRadio}
S.~{Sardellitti}, G.~{Scutari}, and S.~{Barbarossa}, ``Joint optimization of
  radio and computational resources for multicell mobile-edge computing,''
  \emph{IEEE Trans. Signal Inf. Process. over Netw.}, vol.~1, no.~2, pp.
  89--103, Jun. 2015.

\bibitem{Wang2019SmartResourceAllocation}
J.~{Wang}, L.~{Zhao}, J.~{Liu}, and N.~{Kato}, ``Smart resource allocation for
  mobile edge computing: A deep reinforcement learning approach,'' \emph{Early
  Access in IEEE Trans. Emerg. Topics Comput. DOI: 10.1109/TETC.2019.2902661}.

\bibitem{Miettinen2010Energyefficiencymobile}
A.~P. Miettinen and J.~K. Nurminen, ``Energy efficiency of mobile clients in
  cloud computing.'' \emph{HotCloud}, vol.~10, no. 4-4, pp. 1--7, Jun. 2010.

\bibitem{Mahmoodi2019OptimalJointScheduling}
S.~E. {Mahmoodi}, R.~N. {Uma}, and K.~P. {Subbalakshmi}, ``Optimal joint
  scheduling and cloud offloading for mobile applications,'' \emph{IEEE Trans.
  Cloud Comput.}, vol.~7, no.~2, pp. 301--313, Apr. 2019.

\bibitem{Zhou2020}
Y.~{Zhou}, C.~{Pan}, P.~L. {Yeoh}, K.~{Wang}, M.~{Elkashlan}, B.~{Vucetic}, and
  Y.~{Li}, ``Secure communications for {UAV}-enabled mobile edge computing
  systems,'' \emph{IEEE Trans. Commun.}, vol.~68, no.~1, pp. 376--388, Oct.
  2020.

\bibitem{Wang2020}
J.~{Wang}, H.~{Yang}, M.~{Cheng}, J.~{Wang}, M.~{Lin}, and J.~{Wang}, ``Joint
  optimization of offloading and resources allocation in secure mobile edge
  computing systems,'' \emph{IEEE Trans. Veh. Technol.}, vol.~69, no.~8, pp.
  8843--8854, Aug. 2020.

\bibitem{Wu2020}
W.~{Wu}, F.~{Zhou}, R.~Q. {Hu}, and B.~{Wang}, ``Energy-efficient resource
  allocation for secure {NOMA}-enabled mobile edge computing networks,''
  \emph{IEEE Trans. Commun.}, vol.~68, no.~1, pp. 493--505, Jan. 2020.

\bibitem{He2020}
X.~{He}, R.~{Jin}, and H.~{Dai}, ``Physical-layer assisted secure offloading in
  mobile-edge computing,'' \emph{IEEE Trans. Wireless Commun.}, vol.~19, no.~6,
  pp. 4054--4066, Jun. 2020.

\bibitem{Goel2008GuaranteeingSecrecyusing}
S.~{Goel} and R.~{Negi}, ``Guaranteeing secrecy using artificial noise,''
  \emph{IEEE Trans. Wireless Commun.}, vol.~7, no.~6, pp. 2180--2189, Jun.
  2008.

\bibitem{Liu2017SecrecyCapacityAnalysis}
Y.~Liu, H.~H. Chen, and L.~Wang, ``Secrecy capacity analysis of artificial
  noisy {MIMO} channels--{An} approach based on ordered eigenvalues of
  {Wishart} matrices,'' \emph{IEEE Trans. Inf. Forensics Security}, vol.~12,
  no.~3, pp. 617--630, Mar. 2017.

\bibitem{Liu2019ArtificialNoisyMIMO}
Y.~{Liu}, H.~{Chen}, L.~{Wang}, and W.~{Meng}, ``Artificial noisy {MIMO}
  systems under correlated scattering {Rayleigh} fading—{A} physical layer
  security approach,'' \emph{IEEE Syst. J.}, vol.~14, no.~2, pp. 2121--2132,
  Jun. 2020.

\bibitem{Liao2011QoSBasedTransmit}
W.~{Liao}, T.~{Chang}, W.~{Ma}, and C.~{Chi}, ``{QoS}-based transmit
  beamforming in the presence of eavesdroppers: An optimized
  artificial-noise-aided approach,'' \emph{IEEE Trans. Signal Process.},
  vol.~59, no.~3, pp. 1202--1216, Mar. 2011.

\bibitem{3gpp}
``Technical specification group radio access network; evolved universal
  terrestrial radio access {(E-UTRA)} and evolved universal terrestrial radio
  access network {(E-UTRAN)}; overall description; stage 2 (v15.1.0, release
  15),'' no. 3GPP, Tech. Sepc. 36.300, 3 2018, rev. 1.

\bibitem{Yan2017}
Z.~{Yan}, X.~{Li}, M.~{Wang}, and A.~V. {Vasilakos}, ``Flexible data access
  control based on trust and reputation in cloud computing,'' \emph{IEEE Trans.
  on Cloud Comput.}, vol.~5, no.~3, pp. 485--498, Jul. 2017.

\bibitem{Patel2005SimulationRayleighfaded}
C.~S. {Patel}, G.~L. {Stuber}, and T.~G. {Pratt}, ``Simulation of
  {Rayleigh-faded} mobile-to-mobile communication channels,'' \emph{IEEE Trans.
  Commun.}, vol.~53, no.~11, pp. 1876--1884, Nov. 2005.

\bibitem{3gpp2}
``3rd generation partnership project; technical specification group radio
  access network; evolved universal terrestrial radio access ({E-UTRA});
  physical channels and modulation ({R}elease 13),'' no. 3GPP, Tech. Sepc.
  36.211, 6 2016, rev. 13.2.0.

\bibitem{He2013}
B.~{He} and X.~{Zhou}, ``Secure on-off transmission design with channel
  estimation errors,'' \emph{IEEE Trans. Inf. Forensics Security}, vol.~8,
  no.~12, pp. 1923--1936, Dec. 2013.

\bibitem{Li2018}
B.~{Li}, Z.~{Fei}, Z.~{Chu}, F.~{Zhou}, K.~{Wong}, and P.~{Xiao}, ``Robust
  chance-constrained secure transmission for cognitive satellite–terrestrial
  networks,'' \emph{IEEE Trans. Veh. Technol.}, vol.~67, no.~5, pp. 4208--4219,
  Jan. 2018.

\bibitem{Munoz2015}
O.~{Muñoz}, A.~{Pascual-Iserte}, and J.~{Vidal}, ``Optimization of radio and
  computational resources for energy efficiency in latency-constrained
  application offloading,'' \emph{IEEE Trans. Veh. Technol.}, vol.~64, no.~10,
  pp. 4738--4755, Oct. 2015.

\bibitem{Ng2011}
D.~W.~K. {Ng}, E.~S. {Lo}, and R.~{Schober}, ``Secure resource allocation and
  scheduling for {OFDMA} decode-and-forward relay networks,'' \emph{IEEE Trans.
  Wireless Commun.}, vol.~10, no.~10, pp. 3528--3540, Oct. 2011.

\bibitem{Sloane1981}
N.~{Sloane}, ``Tables of sphere packings and spherical codes,'' \emph{IEEE
  Trans. Inf. Theory}, vol.~27, no.~3, pp. 327--338, May 1981.

\bibitem{Raz2002}
R.~Raz, ``On the complexity of matrix product,'' in \emph{2002 Proceedings of
  the thiry-fourth annual ACM symposium on Theory of computing}.\hskip 1em plus
  0.5em minus 0.4em\relax New York, USA: ACM, 2002, pp. 144--151.

\bibitem{3gpp3}
``3rd generation partnership project; technical specification group radio
  access network; {V2X} services based on {NR}; user equipment ({UE}) radio
  transmission and reception; ({R}elease 16),'' no. 3GPP, Tech. Sepc. 38.886, 6
  2020, rev. 16.1.0.

\bibitem{Molina-Masegosa2017LTEVSidelink}
R.~{Molina-Masegosa} and J.~{Gozalvez}, ``{LTE-V} for sidelink {5G V2X}
  vehicular communications: {A} new {5G} technology for short-range
  vehicle-to-everything communications,'' \emph{IEEE Veh. Technol. Mag.},
  vol.~12, no.~4, pp. 30--39, Dec. 2017.

\bibitem{Taubman2000Highperformancescalable}
D.~{Taubman}, ``High performance scalable image compression with ebcot,''
  \emph{IEEE Trans. Image Process.}, vol.~9, no.~7, pp. 1158--1170, Jul. 2000.

\bibitem{Shu2016}
Z.~Shu, J.~Wan, D.~Li, J.~Lin, A.~V. Vasilakos, and M.~Imran, ``Security in
  software-defined networking: {T}hreats and countermeasures,'' \emph{Mobile
  Networks and Applications}, vol.~21, no.~5, pp. 764--776, Jan. 2016.

\bibitem{Parvez2018}
I.~{Parvez}, A.~{Rahmati}, I.~{Guvenc}, A.~I. {Sarwat}, and H.~{Dai}, ``A
  survey on low latency towards {5G}: {RAN}, core network and caching
  solutions,'' \emph{IEEE Commun. Surveys Tuts.}, vol.~20, no.~4, pp.
  3098--3130, Fourthquarter 2018.

\bibitem{Shin2003}
H.~Shin and J.~H. Lee, ``Capacity of multiple-antenna fading channels: spatial
  fading correlation, double scattering, and keyhole,'' \emph{IEEE Trans. Inf.
  Theory}, vol.~49, no.~10, pp. 2636--2647, Oct. 2003.

\bibitem{Chiani2010}
M.~{Chiani}, M.~Z. {Win}, and H.~{Shin}, ``{MIMO} networks: The effects of
  interference,'' \emph{IEEE Trans. Inf. Theory}, vol.~56, no.~1, pp. 336--349,
  Jan. 2010.

\bibitem{McKay2005}
M.~R. {McKay} and I.~B. {Collings}, ``General capacity bounds for spatially
  correlated {Rician MIMO} channels,'' \emph{IEEE Trans. Inf. Theory}, vol.~51,
  no.~9, pp. 3121--3145, Sep. 2005.

\bibitem{Salo2006}
J.~{Salo}, F.~{Mikas}, and P.~{Vainikainen}, ``An upper bound on the ergodic
  mutual information in {Rician fading MIMO} channels,'' \emph{IEEE Trans.
  Wireless Commun.}, vol.~5, no.~6, pp. 1415--1421, Jun. 2006.

\bibitem{Gao1998TheoreticalreliabilityMMSE}
H.~{Gao}, P.~J. {Smith}, and M.~V. {Clark}, ``Theoretical reliability of {MMSE}
  linear diversity combining in {Rayleigh-fading} additive interference
  channels,'' \emph{IEEE Trans. Commun.}, vol.~46, no.~5, pp. 666--672, May
  1998.

\end{thebibliography}

\vfill

\end{document}